# Toward Personalized Medicine in Connectomic Deep Brain Stimulation

**Authors:** Barbara Hollunder[1,2,3*], Nanditha Rajamani[1], Shan H. Siddiqi[4,5], Carsten Finke[1,2,3], Andrea A. Kühn[1,2,3,6], Helen S. Mayberg[7], Michael D. Fox[4,8], Clemens Neudorfer[1,4], Andreas Horn[1,2,4,8]

*Corresponding author

**Affiliations**

1. Department of Neurology, Charité – Universitätsmedizin Berlin, Berlin, Germany
2. Einstein Center for Neurosciences Berlin, Charité – Universitätsmedizin Berlin, Berlin, Germany
3. Berlin School of Mind and Brain, Humboldt-Universität zu Berlin, Berlin, Germany
4. Center for Brain Circuit Therapeutics, Brigham & Women's Hospital, Boston, MA, USA
5. Department of Psychiatry, Harvard Medical School, Boston, MA, USA
6. NeuroCure Cluster of Excellence, Charité – Universitätsmedizin Berlin, Berlin, Germany
7. Nash Family Center for Advanced Circuit Therapeutics, Icahn School of Medicine at Mount Sinai, New York, NY, USA
8. Department of Neurosurgery, Massachusetts General Hospital, Harvard Medical School, Boston, MA, USA

**Corresponding author**

Barbara Hollunder

Movement Disorders and Neuromodulation Unit, Department of Neurology,

Charité – Universitätsmedizin Berlin, Charitéplatz 1, 10117 Berlin, Germany

barbara.hollunder@charite.de

**Word counts**

Number of words in the abstract: 200

Number of words in the main text: 6000

Number of Figures: 7 (plus graphical abstract)

# List of abbreviations

| | |
|---|---|
| ALIC | Anterior Limb of the Internal Capsule |
| BOLD | Blood Oxygenation Level Dependent |
| DBS | Deep Brain Stimulation |
| ET | Essential Tremor |
| GPi | Internal Pallidum |
| MDD | Major Depressive Disorder |
| OCD | Obsessive Compulsive Disorder |
| OFC | Orbitofrontal Cortex |
| PD | Parkinson's Disease |
| RDoC | Research Domain Criteria |
| rs-fMRI | resting state-functional Magnetic Resonance Imaging |
| STN | Subthalamic Nucleus |

## Abstract

At the group-level, deep brain stimulation leads to significant therapeutic benefit in a multitude of neurological and neuropsychiatric disorders. At the single-patient level, however, symptoms may sometimes persist despite "optimal" electrode placement at established treatment coordinates. This may be partly explained by limitations of disease-centric strategies that are unable to account for *heterogeneous* phenotypes and comorbidities observed in clinical practice. Instead, tailoring electrode placement and programming to individual patients' symptom profiles may increase the fraction of top responding patients. Here, we propose a three-step, circuit-based framework that aims to develop patient-specific treatment targets that address the unique symptom constellation prevalent in each patient. First, we describe how a *symptom network target library* could be established by mapping beneficial or undesirable DBS effects to distinct circuits based on (retrospective) group-level data. Second, we suggest ways of matching the resulting symptom networks to circuits defined in the individual patient (*template matching*). Third, we introduce *network blending* as a strategy to calculate optimal stimulation targets and parameters by selecting and weighting a set of symptom-specific networks based on the symptom profile and subjective priorities of the individual patient. We integrate the approach with published literature and conclude by discussing limitations and future challenges.

## Keywords

Circuitopathies, connectomics, deep brain stimulation (DBS), neuromodulation, personalized medicine, Research Domain Criteria.

# Graphical abstract

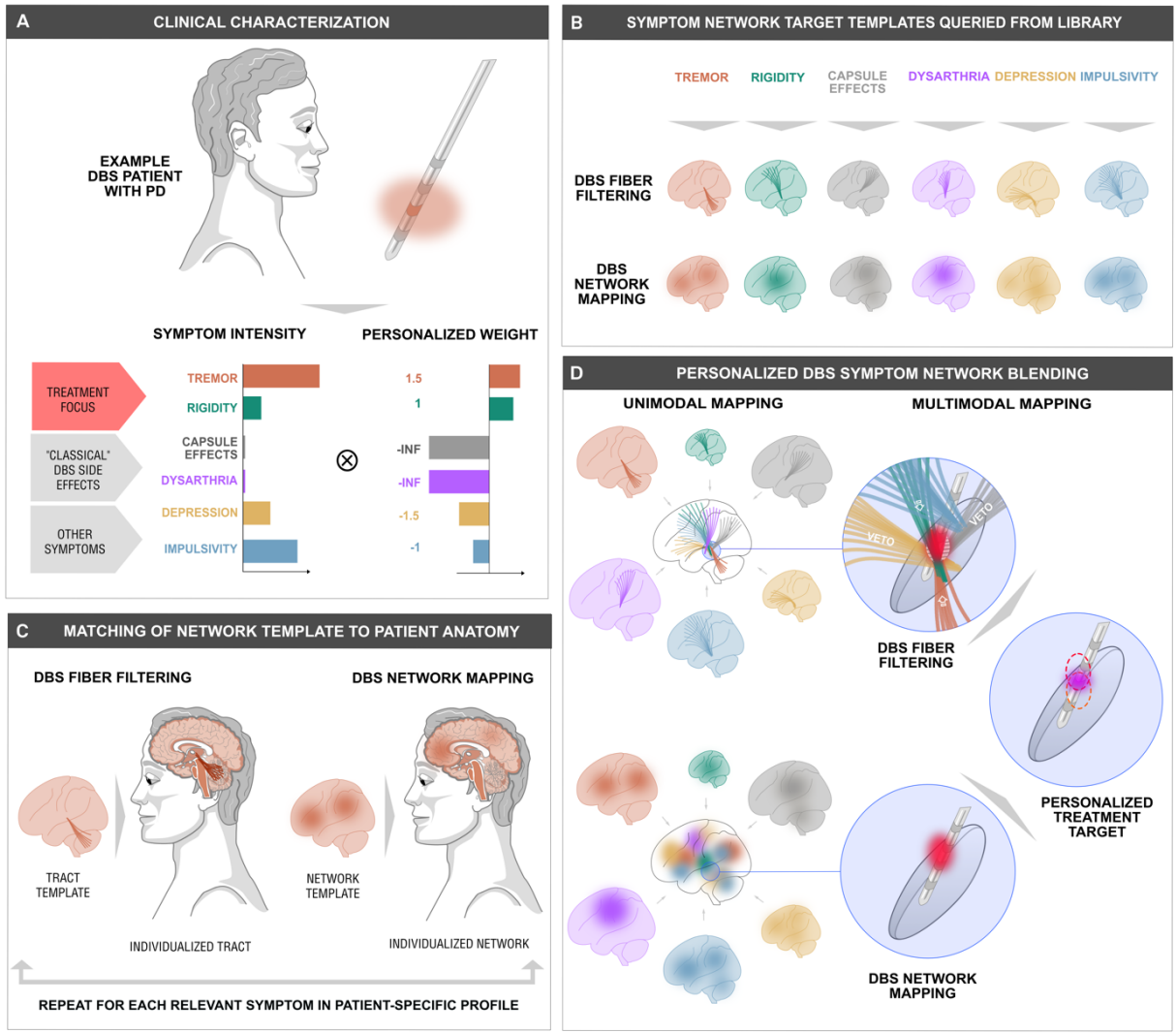

# 1. Introduction

The advent of deep brain stimulation (DBS) has revolutionized neurological therapies and provided a unique window into the pathophysiology of circuit dysfunction (Gardner, 2013; Priori, 2015). Along with technological and methodological advances the field has paved the path to over 200.000 DBS surgeries across a broad spectrum of neurological and psychiatric diseases (Vedam-Mai et al., 2021). Apart from U.S. Food and Drug Administration approvals for Parkinson's disease (PD), essential tremor (ET), and epilepsy, DBS has received humanitarian device exemption for dystonia and obsessive-compulsive disorder (OCD) (Lee et al., 2019). Among the many investigational applications, examples are intractable major depressive disorder (MDD) (Dougherty et al., 2015; Holtzheimer et al., 2017; Mayberg et al., 2005), Alzheimer's disease (Kuhn et al., 2015; Laxton et al., 2010), chronic pain syndromes (Farrell et al., 2018), schizophrenia (Corripio et al., 2020; Wang et al., 2020), and minimally conscious states (Chudy et al., 2018; Yamamoto et al., 2005).

While DBS has reached standard-of-care status in various movement disorders (Fasano et al., 2015; Limousin and Foltynie, 2019; Moro et al., 2017), outcome variability is more pronounced in the neuropsychiatric domain (Alonso et al., 2015; Sullivan et al., 2021; Zhou et al., 2018). Notwithstanding the underlying pathophysiological correlate, response to DBS is dictated by a multitude of factors. These include differences in preoperative targeting strategies (*across* – but also *within* – surgical target sites), intraoperative electrode placement, stimulation contact and parameter selection, placebo effects, side-effects, or clinical and demographic patient characteristics (Limousin and Foltynie, 2019; Pilitsis et al., 2008). Even in established indications, however, symptom relief can remain absent for some patients despite accurate electrode placement (Okun et al., 2005; Pauls et al., 2017).

This latter observation has raised concerns that strategies developed for the "average" patient and their cardinal symptomatology could, in fact, obscure clinically meaningful variability of DBS effects onto the full symptom spectrum at the individual-patient level (Allawala et al., 2021; Figee and Mayberg, 2021). Indeed, both current DBS research and clinical practice predominantly rely on conditional diagnoses for pooling patients into clusters for which best-practice neuromodulation strategies are continuously developed and revised. In view of symptomatic and pathophysiological *heterogeneity* within disorders (Marquand et al., 2016) along with cross-diagnostic *comorbidity* (Husain, 2017; Merikangas et al., 2015; Plana-Ripoll et al., 2019), the tendency toward a disease-centric "one-fits-all" approach may, however, limit personalized precision medicine in DBS (Casey et al., 2013; Cuthbert and Insel, 2013). Notably, even established diagnoses such as PD present with varying clinical phenotypes and the unity of a common pathology is increasingly being questioned (Fearon et al., 2021; Mestre et al., 2021). Moreover, with DBS optimization strategies often focusing on a cardinal set of symptoms for a given disorder, secondary symptomatology may become unmasked and undermine full recovery. For instance, PD patients often benefit from DBS to the subthalamic nucleus (STN) or internal pallidum (GPi) in terms of improved tremor, bradykinesia, rigidity, or motor fluctuations. Instead, symptoms pertaining to speech, affect, cognition – and even to the full range of motor features (e.g., freezing of gait) – may persist or even deteriorate following surgery (Rodriguez-Oroz et al., 2012).

By contrast, an interventional framework of higher adaptability to phenotypical heterogeneity could target an individual's biosignature (or *biotype*) in more granular fashion to enable reliable predictions of single-patient outcome within a comprehensive symptom range (Fernandes et al., 2017; Perlis, 2011; Strimbu and Tavel, 2010). Neuroimaging-based markers have shown particular clinical utility in treatment selection for individuals (Barcia et al., 2019; Drysdale et al., 2017; Dunlop et al., 2017; Kelley et al., 2021; Korgaonkar et al., 2015; McGrath et al., 2013). To define neuroimaging biomarkers for personalized DBS it is worth closely examining its assumed mechanisms of action. DBS emits weak high-frequency pulses of electrical current via electrodes surgically implanted into subcortical white or grey matter targets (Jakobs et al., 2019). Increasingly, researchers focus on widespread DBS network effects above and beyond the focal target itself (Alhourani et al., 2015; Ashkan et al., 2017; Horn and Fox, 2020; Lozano and Lipsman, 2013). Subsequently, the degree of modulation of specific networks has allowed to forecast symptom improvements across DBS cohorts and centers (Horn et al., 2019, 2017; Li et al., 2021, 2020; Okromelidze et al., 2020; Sweet et al., 2020). Establishing a comprehensive library (or atlas) of networks that, when modulated, clearly associate with distinct symptoms thus represents a valuable opportunity to personalize neuromodulation.

This rationale aligns with the agenda of the Research Domain Criteria (RDoC) initiative of the U.S. National Institute of Mental Health, which sees the future of precision medicine in treating neurological and neuropsychiatric conditions as *"disorders of the human connectome"* (or *circuitopathies*) (Gordon, 2016; Insel et al., 2010). The RDoC framework assumes that observable behavior along dimensional symptom axes would map onto distinct circuit dysfunctions (Cuthbert, 2014; Cuthbert and Insel, 2013; Insel, 2014). The latter may be altered in functionally selective ways by means of neuromodulation at any given network node (Crocker et al., 2013; Fox et al., 2014; Horn and Fox, 2020; Insel et al., 2010; Shephard et al., 2021). Going forward, we will use the term of *"symptom network"* to describe a specific network associated with a particular symptom cluster (introduced as *"symptomatotopy"* by Lozano and Lipsman (2013)). When referencing the impact of neuromodulation onto these networks, we will refer to *"symptom network targets"*.

Individual re-combinations and degrees of importance of these symptom networks could explain phenotypical variability across patients. Beyond PD, symptom networks have been defined, for instance, in anxiety and MDD (Drysdale et al., 2017; Liang et al., 2020; Price et al., 2017; Wager and Woo, 2017; Williams, 2017), OCD (Harrison et al., 2013; Mataix-Cols et al., 2004; Thorsen et al., 2018), or psychotic disorders (Clementz et al., 2016; Ivleva et al., 2017) – some of which, however, failed to replicate (Dinga et al., 2019). At the same time, symptom networks could be *shared across* disorders along dimensional symptom axes (Husain, 2017). Compulsivity, reward processing, or inhibition are examples of functional dimensions which may each rely on uniform network underpinnings but contribute to the clinical presentations of *different* disorders (Gillan et al., 2017, 2016; Lansdall et al., 2017; Nusslock and Alloy, 2017; Robbins et al., 2019; van den Heuvel et al., 2016; Whitton et al., 2015; Yücel et al., 2019). Tailoring DBS to a *combination* or *blend* of symptom networks could thus augment disease-specific electrode implantation and programming strategies through added levels of precision and facilitate improvement in unique phenotypes (Figee and Mayberg, 2021; Horn and Fox, 2020). This strategy also links with recent developments in DBS technology – such as directional leads with an option for independent stimulation contact selection – which invite a more modular approach to treating symptom clusters.

Building on concepts like these, here, we intend to formalize an emerging DBS personalization framework by <u>multidimensional symptom profiling via connectomics</u> as grounds for future development. Our manuscript should be seen as a *blueprint* or whitepaper which comprises several scientifically established aspects but also others that remain to be worked out. The proposed approach consists of three consecutive steps which will structure this manuscript, including i) creating group-level *symptom network target libraries*, ii) translating these network targets into patient space based on unique connectivity profiles *(template matching)*, and iii) selecting and weighting individualized networks according to the symptom spectrum of specific patients to synthesize <u>personalized DBS targets</u> *(network blending)*. Ultimately, the framework aims to refine surgical planning (i.e., electrode placement *within* established targets) and postoperative stimulation parameter programming to maximize DBS outcome while avoiding undesirable side-effects.

## 2. Toward a library of symptom networks

### *2.1. "Bottom-up" definition of symptom networks*

Before addressing our main aim of empirically defining a *symptom network target library* using neuromodulation, we touch upon other ways of matching brain networks to corresponding symptoms. Attributing function to specific circuits is a fundamental goal of neuroscience and has been pursued for centuries (Eickhoff et al., 2018).

A classic example of clustering brain function has been to segregate striatal loops and their parallel – but cross-communicating (Aoki et al., 2019; Guthrie et al., 2013; Kolomiets et al., 2001) – interconnections between cortex, basal ganglia, thalamus, and brainstem (Alexander et al., 1986; Alexander and Crutcher, 1990). These cortico-basal ganglia-thalamo-cortical loops have fundamentally defined treatment concepts of brain disorders by their preference to process blends of *sensorimotor*, *limbic,* or *associative* information (**Fig. 1A, left panel**). Domain-specific *"plus"* or *"minus"* symptoms are seen as consequences of dysfunction of respective loops. Namely, authors established parallels between motor *"plus"* symptoms such as dyskinesia (motor) with premature responding (associative) or mania/impulsivity (limbic) as results of hyperfunction or "overmodulation" of the according loop (Volkmann et al., 2010). Similarly, on the *"minus"* side, akinesia (motor) was set into parallel with reduced psychomotor speed (associative), or depression/apathy (limbic) following hypofunction or "undermodulation" (**Fig. 1A, right panel)**.

Further extending this concept of parallel striatal loops, Swanson proposed to generalize the functional description of "striatum" and "pallidum" to (all) remaining brain nuclei, which – based on their connections, neurotransmitter types and embryological development – would implicate network components with "striatal" or "pallidal" functional roles beyond the one involving the classical dorsal/ventral striatum (Swanson, 2003, 2000). Across all cortical loops, these roles would involve a cascading projection of excitation, inhibition, and disinhibition (**Fig. 1B**). Following this logic, Swanson identified several circuits, each of which features a cortical region and regions functionally resembling a "striatum" and a "pallidum". The hippocampo-septo-hypothalamic loop, for instance, comprises hippocampal cortical sites that exert excitatory effects on the lateral septal complex ("striatum"), which in turn inhibits the medial septal nucleus and nucleus of the diagonal band ("pallidum"). Since septal nucleus and diagonal band feature inhibitory projections, the net effect exerted by both structures would yield disinhibition, similarly to the cortico-striato-pallidal loop well defined in the dorsal striatum.

Once fully characterized, this framework could serve as a "bottom up" library of dedicated functional networks in patients undergoing DBS to guide the alteration of distinct symptoms (**Fig. 1B**). Conceivably, *sub*domains of proposed loops could be impacted in *multiple* ways, leading to different symptoms (e.g., tremor, bradykinesia, and dyskinesia as subnetworks of the dorsal striatum motor loop).

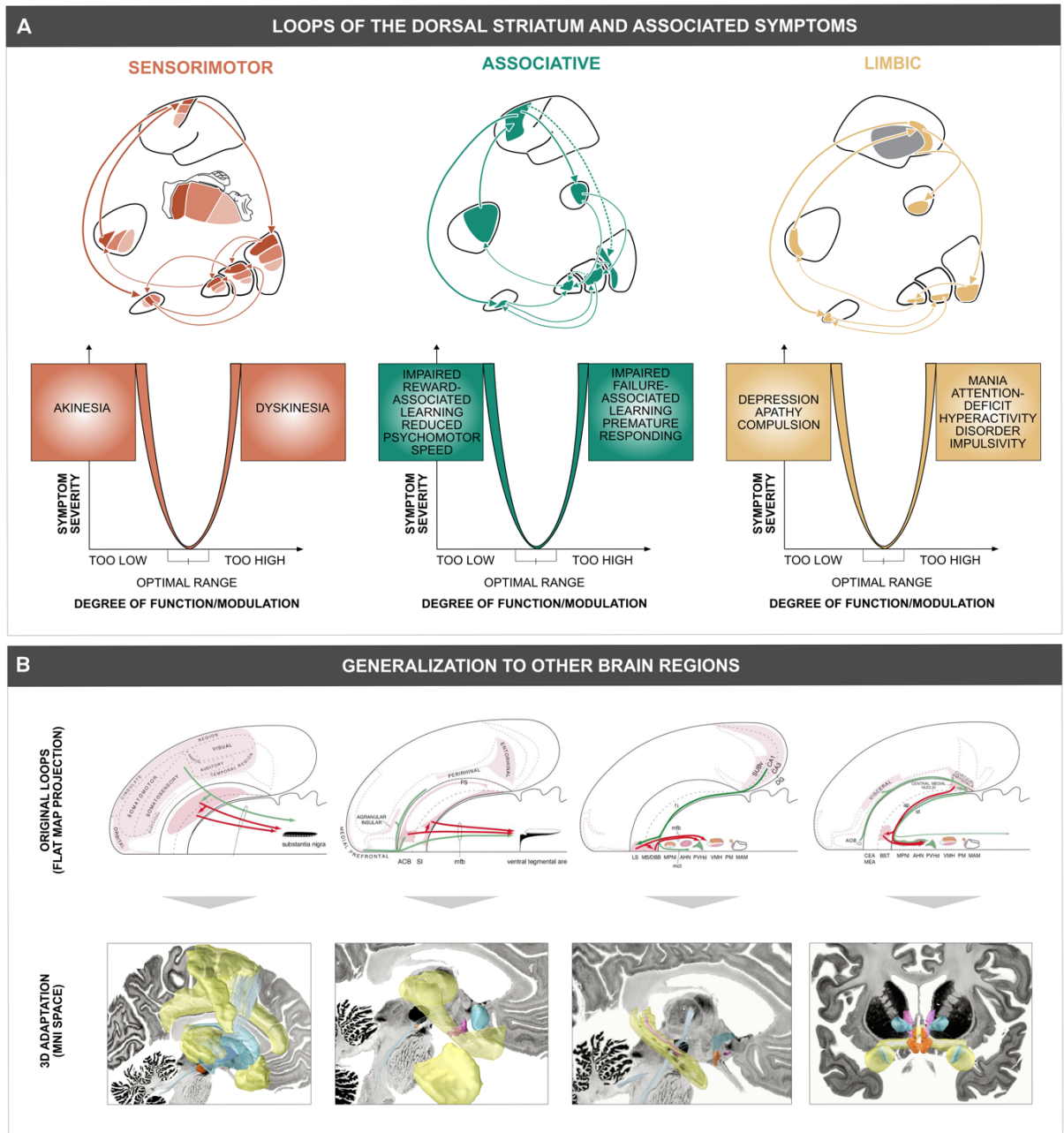

**Fig. 1:** Cortico-striatal loops. **(A)** Classical segregation of the cortico-basal-ganglia-thalamocortical loops along with associated *"plus"* and *"minus"* symptoms as inspired by Volkmann et al. (2010). Symptoms are thought to be induced through "over- or undermodulation" or hyper- or hypofunction. **(B)** Swanson's extension of the striatal and pallidal function to all regions of cortex and brain nuclei. We reproduced four loops as defined by Swanson (2003) on flat-map projections (i.e., cortico-striato-nigral, cortico-striato-tegmental, hippocampo-septo-hypothalamic, and cortico-central/medial-bed nucleus projections) in 3D on the bottom. The BigBrain template is shown as a backdrop (Amunts et al., 2013). Each loop comprises tripartite projections with cortical (excitatory), striatal (inhibitory) and pallidal (disinhibitory) roles to regions within a behavioral control column of motivated behavior. For instance, in the third loop shown, the lateral septal complex takes a "striatal" role, while the medial septal nucleus und nucleus of the diagonal band take the "pallidal" role. Note that the first loop would comprise both motor and associative loops of panel A – suggesting that Swanson's loops should be further segregated. Across all four loops, yellow regions indicate cortical, blue striatal, purple pallidal, and orange hypothalamic regions, while fiber connections are displayed in grey. Flat map projections in panel B reproduced, with permission, from original work (Swanson, 2000). *Abbreviations:* ACB, nucleus accumbens; AHN, anterior hypothalamic nucleus; AOB, accessory olfactory bulb; ap, ansa peduncularis; BST, bed nuclei stria terminalis; CA1/3, fields of Ammon's horn; CEA, central amygdalar nucleus; fi, fimbria; FS, striatal fundus; LS, lateral septal complex; MAM, mamillary body; mct, medial corticohypothalamic tract; MEA, medial amygdalar nucleus; mfb, medial forebrain bundle system; MPN1, medial preoptic nucleus, lateral part; MS/DBB, medial septal/diagonal band nuclei; PM, premamillary nuclei; PVHd, descending paraventricular nucleus; SI, substantia innominata; st, stria terminalis; SUBv, ventral subiculum; VMH, ventromedial hypothalamic nucleus.

## 2.2. "Top-down" definition of therapeutic symptom network targets

While the bottom-up approach may provide a basis to tailor surgical interventions towards specific symptom domains, these concepts have not yet informed approaches in humans. In contrast, lesion studies and surgical interventions within deep brain targets have inspired some of the most influential cortico-basal ganglia models as well as the concept of circuitopathies and *symptom networks* (Bergman et al., 1990; Deffains et al., 2016; DeLong, 1990). Neuromodulation studies themselves may constitute a powerful way to probe and characterize changes along specific symptom dimensions across *symptom networks* (Fox et al., 2014; Horn and Fox, 2020; Siddiqi et al., 2021, 2020). DBS may be particularly suited, since i) stimulation sites are highly focal (i.e., specific), and ii) DBS exerts strong and long-lasting effects on symptoms. Specifically, by means of DBS, the acute and focal manipulation of a precisely targeted region that forms part of a network (*cause*) can be linked to a pronounced behavioral effect by their temporal sequence (Etkin, 2019, 2018).

In recent years, symptom network targets have been increasingly investigated on the group level to harness DBS outcome variability: If electrodes in all top-responding patients fell into a particular network, but the same network was *not* targeted in poor-responding patients, researchers were able to infer a clear – and potentially causal – relationship between network modulation and overt effects (Horn, 2019; Horn and Fox, 2020).

As an example from the motor domain, the degree of structural connectivity between electrode sites and the supplementary motor area accounted for *bradykinesia* and *rigidity* improvements in PD patients receiving STN-DBS (Akram et al., 2017). On the contrary, it was possible to explain variance in *tremor* improvements based on the degree of stimulation applied to the cerebello-thalamo-cortical network. Studies in ET patients with DBS to the ventrointermediate thalamic nucleus associated the same network with tremor suppression (Akram et al., 2018; Al-Fatly et al., 2019). Extending this concept, Coenen et al. (2020) associated stimulation of the dentato-rubro-thalamic tract with tremor improvements in ET, PD, multiple sclerosis, or dystonic tremor patients (**Fig. 2A**) – although disease-specific networks also appear to play a role (Tsuboi et al., 2021). Further subdividing this network based on somatotopy, optimal DBS for maximal treatment success in *head* versus *hand tremor* was associated with modulation of *head* versus *hand* networks in primary motor area and cerebellum (Al-Fatly et al., 2019).

Similarly, studies have begun to map specific networks to DBS effects in the *neuropsychiatric* domain. In a prospective OCD trial, DBS to the anteromedial STN improved *cognitive flexibility*, mediated via connectivity to lateral orbitofrontal (OFC), dorsal anterior cingulate, and dorsolateral prefrontal cortices (Tyagi et al., 2019). In contrast, effectiveness of DBS to the anterior limb of the internal capsule (ALIC) in the same six patients was higher in ameliorating *mood* symptoms, underpinned by electrode connectivity to the medial OFC. A similar finding had previously been detected in MDD-DBS (Riva-Posse et al., 2014). A pathophysiological intersection in form of a reward network involving the medial OFC in both OCD (Figee et al., 2011) and MDD (Cheng et al., 2016) may explain why patients affected by either disorder often benefit from DBS to the ALIC (which entertains connections to this circuit) (Figee and Mayberg, 2021).

In the study by Tyagi et al. (2019) – above and beyond their symptom-specific roles – both STN and ALIC targets were *similarly effective* in reducing global obsessive-compulsive behaviors in OCD patients. This implies a range of "optimal" entry nodes to pinpoint a shared therapeutic symptom network. Indeed, recent work by our group confirmed that stimulating both anteromedial STN and ALIC sites modulates a common network associated with effective DBS response. Specifically, this circuitry was characterized in form of a tract-based target in Li et al. (2020) and a volume-based functional network target in Li et al. (2021). The tract-based target has since been validated by other research groups (Baldermann et al., 2021; Smith et al., 2021; van der Vlis et al., 2021).

In accordance with the RDoC framework and as exemplified above for tremor or reward processing, some network targets may further span *across disorders* given phenotypical overlap along a *shared* symptom dimension. Published examples supporting such a transdiagnostic rationale in the neuromodulation context are given in **Fig. 2.** In the future, concepts like these could become increasingly useful to augment disease-specific strategies and encourage knowledge transfer between disorders (Cuthbert, 2014).

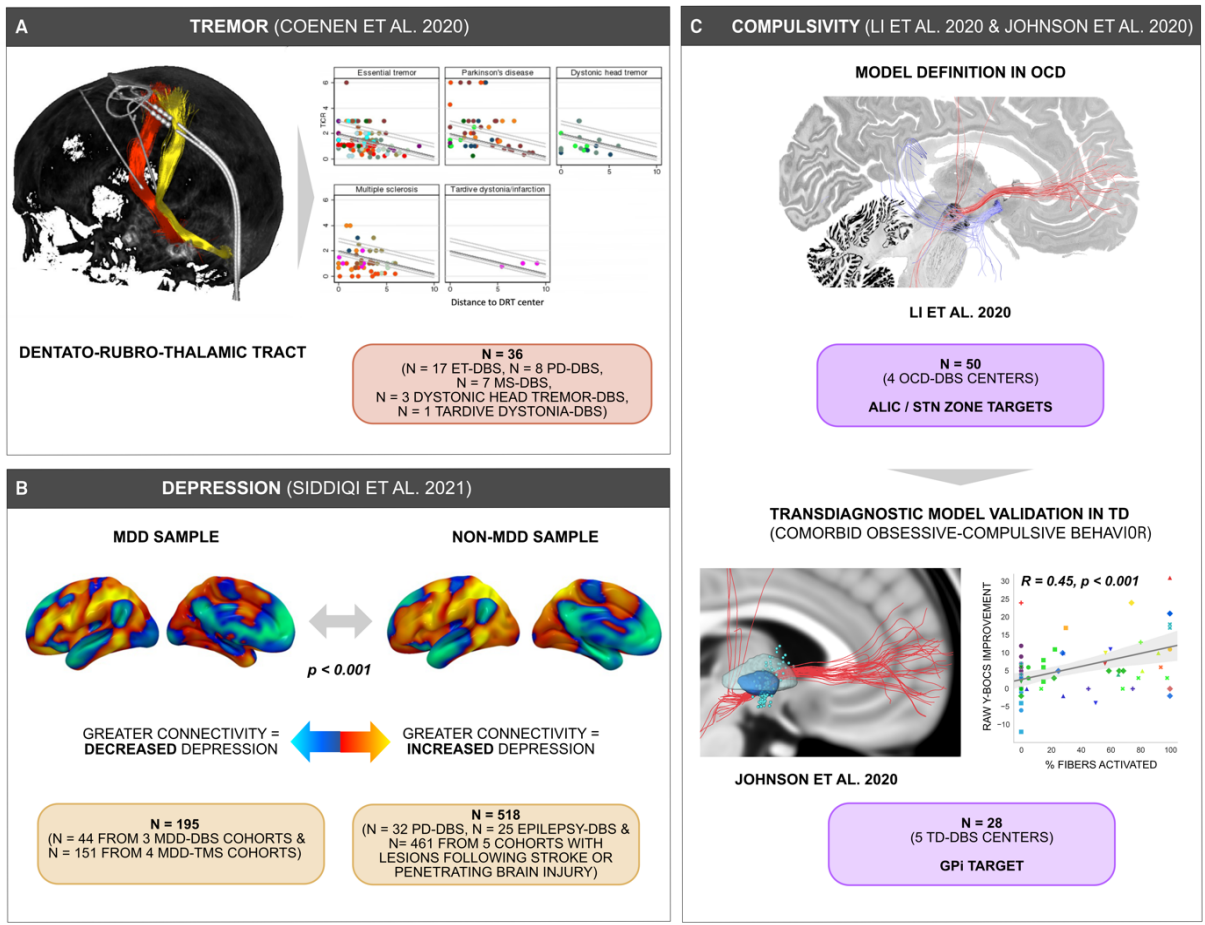

**Fig. 2:** Examples of transdiagnostic symptom network targets. **(A)** In an observational case series, Coenen et al. (2020) probed the dentato-rubro-thalamic tract (DRT) as a common deep brain stimulation (DBS) target for tremor of differential origin (essential tremor [ET], Parkinson's disease [PD], multiple sclerosis [MS], dystonic head tremor, tardive dystonia). The current amplitude required for achieving tremor reduction (tremor improvement per current ratio [TiCR]) of patients who underwent tractography-assisted DBS was significantly ($p < 0.01$) negatively associated with the distance to both DRT center (see panel A) and border (not displayed here). **(B)** Siddiqi et al. (2021) demonstrated brain circuits associated with severity of depressive symptoms to be highly similar ($p < 0.001$) between i) a major depressive disorder (MDD) and ii) a non-MDD cohort, treated via different brain stimulation modalities or experiencing depressive symptoms following lesions of different cause. Specifically, a nearly identical network emerged when integrating a normative high-resolution connectome ($n = 1.000$) with connectivity seeds from i) antidepressant stimulation sites in patients treated for MDD via either DBS or transcranial magnetic stimulation (TMS), or from ii) either depression-inducing or antidepressant stimulation sites of patients treated with DBS for epilepsy or PD, and from depression-inducing lesions caused by penetrating brain injury or stroke. **(C)** Based on the degree of activation of a white-matter pathway that had initially been characterized as unifying tract target in DBS for obsessive-compulsive disorder (OCD) for treating global obsessive-compulsive symptomatology (Li et al., 2020), Johnson et al. (2020b) were able to predict improvements of obsessive-compulsive behavior in Tourette disorder (TD) patients after DBS to the globus pallidus internus (GPi). Importantly, the original tract target had been calculated on data of patients receiving DBS to either anterior limb of the internal capsule (ALIC) or anteromedial subthalamic nucleus (STN) target zones, but not on GPi-DBS data. Figures in panel A adapted from Coenen et al. (2020) under a Creative Commons Attribution 4.0 International license (http://creativecommons.org/licenses/by/4.0/). Figures in panels B and C reproduced, with permission, from Siddiqi et al. (2021) and Baldermann et al. (2021), respectively. *Abbreviations:* Y-BOCS, Yale-Brown Obsessive-Compulsive Scale.

### *2.3. Defining symptom networks based on side-effects*

While symptom network targets can be defined based on intended neuromodulation effects, the same can be achieved for undesired side-effects (Horn and Fox, 2020). In this context, Al-Fatly et al. (2019) associated networks with stimulation-induced ataxia and dysarthria based on a cohort of 33 ET patients. In some cases, side-effects – such as hypomania in STN-DBS for PD (Coenen et al., 2009) – have even inspired novel treatment targets for different diseases such as MDD (Schlaepfer et al., 2013) and OCD (Coenen et al., 2017). Hence, modulating the same symptom network could lead to either desired or unwanted patient-specific effects in dependence of pre-operative symptomatology.

Finally, some findings could potentially inform *"veto"* network-targets (i.e., networks associated with DBS side-effects). On a focal level, reversal of detrimental DBS outcome could be achieved by decreasing the amount of stimulation to territory associated with stimulation-related side-effects (Frankemolle et al., 2010). Extending this focal concept toward circuit-level DBS, specific side-effect networks may be identified that should be avoided by DBS – for instance, such that have been associated with depressive symptoms in PD by Irmen et al. (2020), or with weight changes in patients receiving ALIC-DBS for treatment of OCD/addiction (Baldermann et al., 2019a). Similarly, networks associated with other symptoms, such as slurred speech, impulsivity (Mosley et al., 2020), panic (Elias et al., 2020), apathy, aggression (Yan et al., 2020), dysesthesia, or pain (Cury et al., 2020) should be spared by DBS.

### *2.4. Aggregating a library of symptom network target templates*

The examples above illustrate the powerful use of DBS to subdivide network targets into distinct symptom domains, which could – once defined on group level – be tailored to patients' individual phenotypical profiles. Hence, we propose to identify symptom network targets by relating networks or fiber bundles activated by DBS to a symptom/side-effect dimension in question. For this purpose, we currently dispose of different resources (**info box A, Fig. 3**) and methods (**info box B, Fig. 4**). Ultimately, not only DBS but also other neuromodulation strategies or brain lesions could inform symptom networks, especially for concepts and behavioral traits inaccessible via DBS (e.g., spirituality (Ferguson et al., 2021) or criminal behavior (Darby et al., 2018)). Integrating multiple sources could further strengthen their reliability (Fox et al., 2014; Siddiqi et al., 2021).

By aggregating and continuously refining these templates we envision to establish a *library of symptom network templates* that could serve as a collection of building blocks for personalization. In view of the intention of *personalizing* DBS, establishing a *normative* symptom network target library, as a first step, may appear counter-intuitive. While this first step can be conceived as a means of anchoring networks to function on a group-level, the consecutive two steps aim at individualizing therapy.

**Info box A: Connectomic resources**

Owing to neuroimaging advances, we have become poised to scrutinize *in vivo* models of remote functional networks or white matter pathways activated by electrical stimulation. One frequently used method is diffusion-weighted magnetic resonance imaging (dMRI), which estimates white matter pathways based on the directionality of anisotropic water molecule diffusion along axons (Jeurissen et al., 2019; Maier-Hein et al., 2017). In contrast, resting-state functional magnetic resonance imaging (rs-fMRI) investigates spontaneous blood oxygenation level dependent (BOLD) signal fluctuations in individuals at rest as a function of the hemodynamic response in activated neuron populations (Fox and Raichle, 2007). The intention is to identify conjointly (de)activating brain areas.

Both techniques can be used to estimate connectivity information either directly from a patient (*individualized* connectivity) (Akram et al., 2017; Baldermann et al., 2019b; Riva-Posse et al., 2018; Tyagi et al., 2019; Vanegas-Arroyave et al., 2016) or indirectly via *normative* connectomes acquired from an independent group of participants (Al-Fatly et al., 2019; Baldermann et al., 2019b; Calabrese et al., 2015; Cash et al., 2019; Horn et al., 2017; Irmen et al., 2020; Li et al., 2020; Riva-Posse et al., 2014; Weigand et al., 2018). Normative connectomes represent wiring atlases of the average healthy (or disease-specific) human brain that allow for robust global-scale insights (Wang et al., 2021), but lack information about individual anatomical/functional variance (Braga and Buckner, 2017; Finn et al., 2015; Gordon et al., 2017b; Kong et al., 2019; Llera et al., 2019). Prospectively, we see utility in both normative and individualized resources, the former to establish symptom network libraries, the latter to fine-tune them in individuals (**Fig. 3**).

Given that both dMRI and rs-fMRI are inherently limited by factors such as low test-retest reliability and susceptibility to false-positive tracts/voxels (Jeurissen et al., 2019; Maier-Hein et al., 2017), efforts have been directed towards alternatives to provide additional anatomical detail. Histological templates (Alho et al., 2020), nonhuman tracer data (Haynes and Haber, 2013; Rohlfing et al., 2012), normative atlases curated by expert anatomists (Petersen et al., 2019), or text-book anatomy could be used in conjunction with existing strategies (Li et al., 2020; Treu et al., 2020). The major challenge in their application, however, is the accurate translation into patient space.

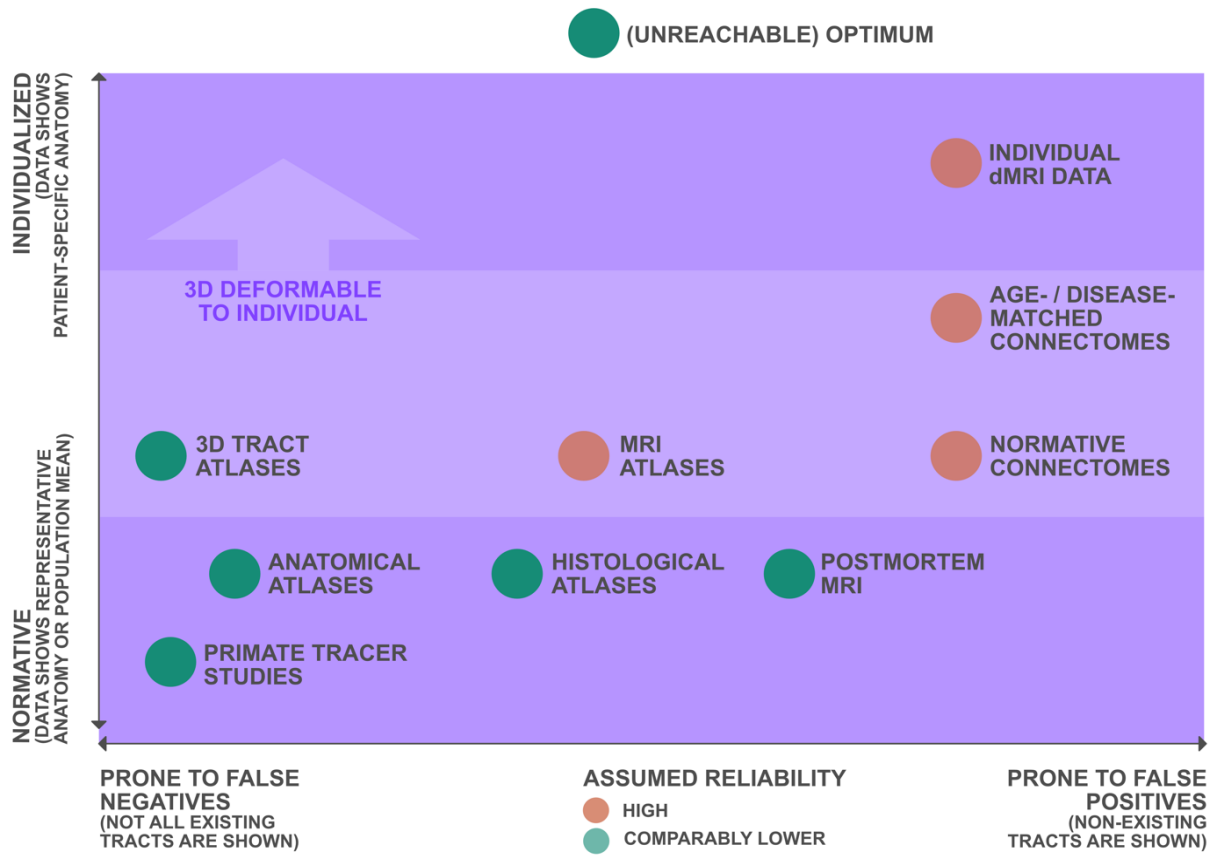

**Fig. 3:** Characteristics of tract-based connectomic resources for deep brain stimulation (DBS) network modeling. Available resources based on structural anatomy are compared as a function of their assumed reliability, their proneness to misrepresenting (i.e., over- or underrepresenting) empirically existing tracts, and how well they match individual anatomy. Resources that are defined in normative space could, in theory, be related to patient-specific anatomy (see bright box with arrow). Importantly, trade-offs are inherent to all these resources and one single, purely advantageous optimum resource is likely unachievable. Also note that this graph does not represent empirical data but is intended as a comparative and non-exhaustive graphical overview. *Abbreviations:* MRI, magnetic resonance imaging; dMRI, diffusion-weighted magnetic resonance imaging.

**Info box B: Methodological primer to define network targets using DBS**

To establish symptom network targets, connectomic DBS models can be linked with changes (i.e., pre- to postoperative/ON- versus OFF-DBS) in specific symptoms (Al-Fatly et al., 2019; Choi et al., 2015; Coenen et al., 2011; Horn et al., 2019, 2017; Okromelidze et al., 2020), behaviors (De Almeida Marcelino et al., 2019; Lofredi et al., 2021; Neumann et al., 2018), or side-effects (Baldermann et al., 2019a; Cury et al., 2020; Irmen et al., 2020; Mosley et al., 2020). Contributing prospective studies could be based on patient stratification into neurobiologically meaningful subtypes (Bell, 2014; Kapur et al., 2012) or dimensional symptom phenotyping. To ensure sufficiently powered retrospective investigations, the use of large repositories of pooled DBS data is advisable (Deeb et al., 2016; Synofzik et al., 2012).

We have proposed two methodological concepts to calculate DBS-based symptom network target templates which could be stored in a library for future use: i) *DBS fiber filtering* (Baldermann et al., 2019b) and ii) *DBS network mapping* (Horn et al., 2017). Code and graphical user interfaces are openly available within Lead-DBS software (www.lead-dbs.org).

DBS fiber filtering (**Fig. 4B**) aims at answering the question which *structural fiber tracts* account for changes in a certain symptom during electrical stimulation. The resulting tract represents the optimal connectivity profile of DBS electrodes for maximized symptom improvement – or "veto" tracts associated with undesirable side-effects. This *tract-based* method has, for instance, been employed to cross-predict variance in OCD-DBS (Baldermann et al., 2019b; Li et al., 2020), or depressive DBS side-effects in PD (Irmen et al., 2020).

While *DBS fiber filtering* allows to visualize one segregated network link (or *edge*) associated with clinical DBS effects, the method is agnostic to potential *indirect* connections to brain regions (or *nodes*) within widely distributed whole-brain networks. DBS network mapping (**Fig. 4C**) can overcome this limitation by incorporating rs-fMRI data representing BOLD signal changes across wide-spread brain regions. This *voxel-based* approach has been used to create DBS symptom network models in PD (Horn et al., 2019, 2017; Irmen et al., 2020), ET (Al-Fatly et al., 2019), OCD (Baldermann et al., 2019b; Li et al., 2021; Sheth et al., 2013), epilepsy (Middlebrooks et al., 2018), or dystonia (Okromelidze et al., 2020).

> The resulting streamlines or R-maps could be stored in form of three-dimensional objects within a *symptom network target library* (**Fig. 4D**), precisely defined in a standard brain template atlas. Important steps toward replicable, neurocircuit-based stimulation correlates could consist of making initial network target atlases available for validation/falsification. Several studies have already successfully adopted this workflow (Johnson et al., 2020; Mosley et al., 2021; Smith et al., 2021).

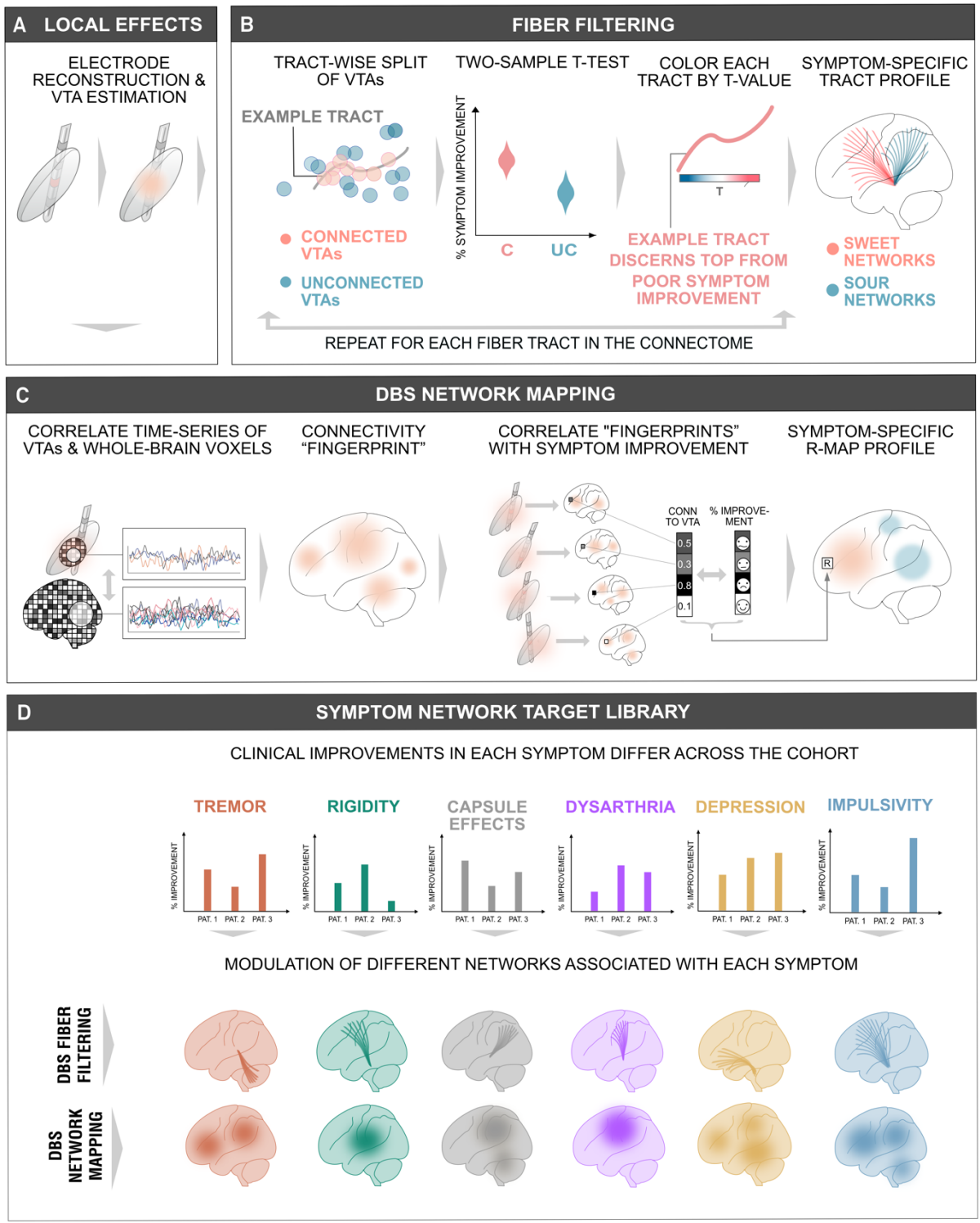

**Fig. 4:** Modeling strategies for deep brain stimulation (DBS) symptom network target templates. **(A)** A network model of DBS effects onto specific clinical outcome parameters can either be estimated on a tract- or voxel-level. In both cases, electrodes are first reconstructed in stereotactic standard space and volumes of tissue activated (VTAs) by the electrical current are estimated. **(B)** In the tract-based approach (Baldermann et al., 2019b), patients are first grouped as a function of whether their VTAs are *connected* (C) or *unconnected* (UC) to a specific fiber tract. Then, the *t*-value resulting from a statistical

comparison between symptom improvement values of both groups is used to color-code the respective tract (here, *red* illustrates tracts associated with stimulation-dependent symptom improvement, while *blue* indicates worsening). This procedure is repeated across the connectome, leading to a symptom-specific model of "optimal" structural electrode connectivity for maximal clinical improvements. *Sweet networks* represent coordinates alongside fiber tracts associated with favorable stimulation outcome, whereas *sour networks* are linked to detrimental DBS effects. **(C)** In the voxel-based approach (Irmen et al., 2020), correlations of mean time-series of voxels within each patient's bilateral VTAs with every remaining whole-brain voxel are performed on a voxel-by-voxel basis. In view of individual differences in localization of active electrode contacts (owing to variability in precise electrode positioning and stimulation parameter programming), this procedure results in one unique functional connectivity (FC) *"fingerprint"* per patient. Voxel-wise correlation of these patient-specific fingerprints with symptom improvements across the patient sample leads to an "optimal" *R-map* model associated with maximal symptom improvement (here, *red* codes for brain regions to which DBS electrodes should ideally connect, while *blue* indicates such areas to which DBS electrodes should optimally be anticorrelated, to ascertain best possible outcome). **(D)** The resulting structural or functional, symptom-specific connectivity models may finally be integrated into a comprehensive, neurocircuit-based taxonomy of DBS effects, which – after thorough replication and refinement – may serve as a neurobiologically meaningful basis for personalization. *Abbreviations:* Conn, connectivity; Pat., patient.

# 3. Template matching: re-discovering symptom networks in individual patients

After defining a library of symptom network target templates, the latter could be re-discovered *within an individual brain* by matching each group-level library entry to intrinsic networks in the respective patient's brain (**Fig. 5**). We refer to this process as *"template matching"*. Specific approaches of realizing exactly this concept (referenced by the same term) have been developed for applications outside of the neuromodulation field (Gordon et al., 2017b, 2017a).

In the DBS context, exploration of the concept and methodological workflows for matching network templates to patient-specific brain anatomy is only beginning to gain traction. One of the few prominent examples is template matching of individualized tracts in MDD-DBS patients to a reference (or "*blueprint*") set of connections that have been termed the *"depression switch"* (Choi et al., 2015). In this line of studies pioneered by the Mayberg group, modulating a set of white matter tracts (forceps minor, cingulum, uncinate fascicle, and fronto-striatal fibers) was associated with optimal improvement of depressive symptoms following DBS of the subcallosal cingulate cortex (Riva-Posse et al., 2014). This set of tracts was defined as a *template* but re-identified using dMRI-based tractography in *individual* patients and prospectively targeted (Riva-Posse et al., 2018).

Similarly, Coenen and colleagues used individualized tractography in MDD patients to identify the ventral tegmental area-projection pathway as a specific tract target (Coenen et al., 2018, 2017; Schlaepfer et al., 2014, 2013), which had been serendipitously defined in a PD patient with hypomania (Coenen et al., 2009). Further, they pioneered prospective targeting (and intraoperative use of diffusion tractography) in ET by stimulating the dentato-rubro-thalamic tract as defined in individual patients (Coenen et al., 2011). **Info box C** contains a primer on methodological avenues toward template matching.

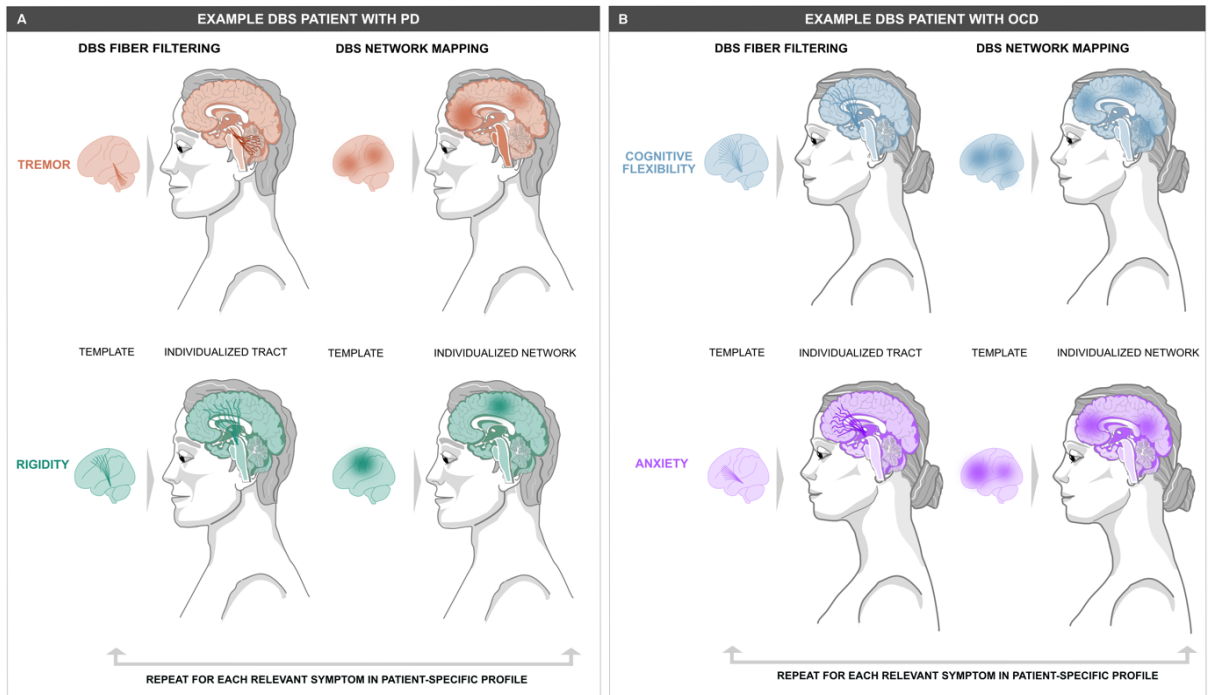

**Fig. 5:** Template matching for application in deep brain stimulation (DBS). Group-average symptom network target templates for DBS with relevance to a patient's unique symptom profile could be queried from the normative library and matched to networks defined in an individual patient. Two example patients are shown, of which one is affected by Parkinson's disease (PD; **panel A**) and the other by obsessive-compulsive disorder (OCD; **panel B**). Depending on the method used to derive the respectively selected templates, multiple strategies are conceivable for template matching, including – *inter alia* – inverse normalization or independent component analysis. While several approaches to template matching have already been probed in other research fields, the precise details of these methods remain to be established and validated in the DBS context.

**Info box C: Methodological primer to match individual networks to symptom templates**

Once a library of symptom-specific tracts is established, one could re-discover the respective tracts in individuals as proposed by the Petersen, Mayberg and Coenen groups. Manual identification of tracts in individual brains or landmark seedings based on template tract definitions could facilitate this process. In contrast, fully data-driven approaches could involve whole-brain tractography in individuals (Reisert et al., 2011) and subsequent tract-wise rating based on spatial similarity of the (coregistered) template tract (e.g., via root mean square distances/k-nearest-neighbor searches across tract-defining points). In this context, automatic tractography and shape-analysis may be useful to re-identify tracts in individuals (Yeh, 2020).

In the functional domain, two candidate approaches can be differentiated: i) scanning functional networks at rest (rs-fMRI) and ii) deriving BOLD signal fluctuations during task performance. In the former, we would aim to find the intrinsic brain network (based on individual rs-fMRI acquired in an individual patient) which best matches the template. Concepts such as independent or principal component analysis have helped clustering intrinsic brain activity into subnetworks (Calhoun et al., 2008; Smith et al., 2012). These subnetworks could be compared to target network templates using distance metrics. Similar approaches for template-matching have applied seed-based connectivity concepts (Gordon et al., 2017a, 2017b). On the other hand, tasks related to the symptom of interest could be investigated via task-based fMRI. For instance, Barcia et al. (2019) used a patient-specific symptom-provocation paradigm to identify treatment networks in OCD. Similarly, Anderson et al. (2011) elucidated patient-specific regions of interest in the motor cortex and superior cerebellum via a finger-tapping task to derive individualized thalamic DBS targets for treatment of ET.

## 4. Network Blending: Toward personalized connectomic DBS

After their re-identification in the individual patient, symptom networks represent valuable resources to adapt DBS targeting and stimulation parameter programming strategies to a patient's unique symptom profile (Horn and Fox, 2020). For a future of individualized DBS, we envision preoperative phenotyping alongside various dimensional symptom axes to inform i) the prevalent *symptom profile* of a specific patient, and ii) the selection of a set of corresponding DBS symptom network targets along with undesirable networks to be avoided by the stimulation (associated with either "classical" DBS side-effects or unwanted symptoms) (**Fig. 6**). The preselected network targets could then iii) be weighted to account for relevant factors beyond empirical symptom intensities, and, finally, iv) be synthesized into a DBS network target tailored to the needs of the individual patient.

In cases where relevant symptom networks are intersecting, networks could be *blended* to identify stimulation sweet spots. While methods of network blending are likely to become more elaborate over time, we propose a simple linear weighting of symptom networks (**Fig. 6**). An optimal electrode implantation coordinate or stimulation setting would maximize stimulation of *relevant* symptom network targets with highest weights while avoiding side-effect networks. In complex cases with non-adjacent symptom networks, current steering via directional electrodes or secondary targets could be suggested. The latter may involve non-conventional trajectories, multiple electrodes or electrodes that reach different regions in one pass. **Info box D** introduces these potential network blending strategies.

Further, we propose a personalized weighting step such that a set of symptoms could form the *treatment focus*, while additional comorbidities could be treated with lesser importance. Moreover, importance of side-effect networks may vary across individuals. While cognitive impairment or affective symptom networks should be avoided in all patients, their avoidance could receive even stronger emphasis in late-stage PD patients that already experience these symptoms. Similarly, beneficial networks could not only be weighted based on symptom severity, but also based on individual preferences. For instance, some patients may experience tremor as a stronger subjective burden than bradykinesia, or vice versa.

Since most symptom libraries will be defined based on conventional DBS target sites (e.g., the STN), we expect coordinate peaks *within* these structures in dependence of the respective symptom profile (Boutet et al., 2021). As a concrete example, based on literature results and unpublished data we anticipate an optimal coordinate for tremor suppression at more dorsal and posterior levels (slightly outside the STN), whereas a maximal bradykinetic effect could reside more ventrally, within STN proper.

Finally, the proposed personalization framework could facilitate harmonization with different forms of neuromodulation, such as other invasive methods (e.g., gamma-knife surgery) or non-invasive forms of brain stimulation, pharmacotherapy, or behavioral interventions (e.g., cognitive-behavioral therapy or neurofeedback) (**Fig. 7**). For instance, invasive and non-invasive brain stimulation modalities could be used to steer shared therapeutic network effects from different vantage points (Fox et al., 2014; Siddiqi et al., 2021) or to access symptom-specific circuits (Crocker et al., 2013; Shephard et al., 2021). Hence, multi-modal strategies such as brain stimulation combined with pharmacotherapy (Sharma et al., 2012), psychological/behavioral interventions (Görmezoğlu et al., 2020; Mantione et al., 2014; Tyagi et al., 2019), light therapy (Weigand et al., 2021), or physical exercise (Miterko et al., 2021) could be of use for optimally tailoring interventions to symptom heterogeneity by balancing modality-specific effectiveness.

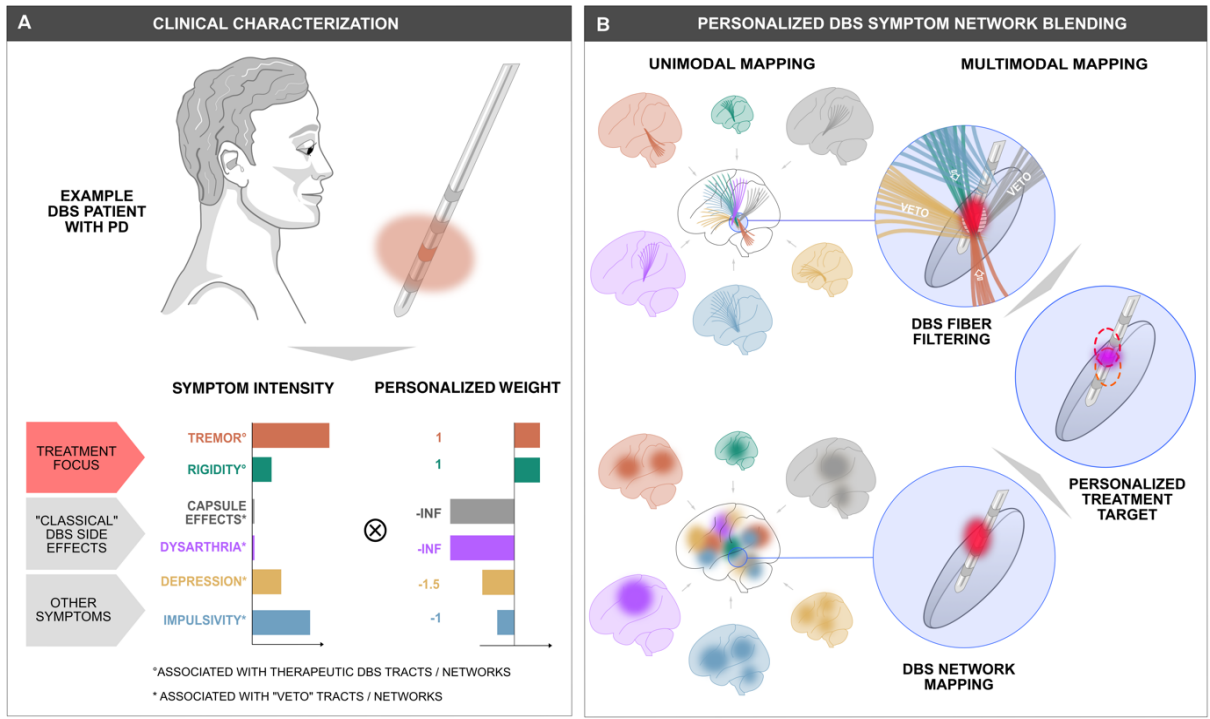

**Fig. 6:** Personalized deep brain stimulation (DBS) symptom network blending. (**A**) In a first step, the prevalent symptom profile of a patient is characterized along different dimensional axes as a function of their respectively experienced intensities. Based on the resulting clinical symptom profile, a corresponding set of therapeutic DBS circuit targets, as well as "veto" networks could be queried from the symptom network template library. Although several symptoms will form the treatment focus, "classical" DBS side effects may be considered, as well as other symptoms that may be present in the patient prior to surgery. In addition, a personalized weighting step could allow to further fine-tune the patient's treatment profile. While side-effects would be attempted to be avoided as much as possible in all patients, a patient's personal circumstances, preferences and needs could also be incorporated at this stage. (**B**) In a second step, a personalized stimulation coordinate most beneficial for the unique phenotype of this patient could be achieved by weighting, synthesizing, and blending the selected symptom network target templates and "vetoing" side-effect targets (which have ideally already been matched with individualized connectivity of the patient in question during the *template matching* step). This "weighting-and-blending" strategy could theoretically be applied to target templates derived via either DBS fiber filtering or via DBS network mapping. Thus, in a final step, optimal stimulation coordinates proposed for each of these modalities separately could be integrated into one most optimal, patient-specific stimulation target. *Abbreviations:* PD, Parkinson's disease.

**Info box D: Methodological primer for DBS network blending**

In view of increasing degrees of freedom in the parameter space, an automated algorithm could represent the most efficient way of suggesting optimal DBS stimulation sites. Symptom intensity weighting could be performed on either a voxel- or a tract-level (depending on the method underlying the selected network target templates). An optimal stimulation coordinate could be based on maximal overlap with fiber tracts/voxels with highest weights. Such an algorithm would optimally also account for other priors of DBS effectiveness to achieve most precise predictions. For instance, patient-specific factors (such as age, disease progression, or medication) may explain additional amounts of variance in clinical outcomes (Cavallieri et al., 2021; Guzick et al., 2020; Shalash et al., 2014).

Crucially, *expert decision making* will be needed to supervise and interpret results in any scenario, even more so if no clear peak stimulation target can be identified. Suggested stimulation sites residing largely outside established treatment targets would need to be disregarded or further investigated in basic research. In some cases, proposed anatomical sites may also be inaccessible via DBS leads. Practically speaking, we see direct benefit of the approach to refine existing target sites (such as subzones within the STN/ALIC), rather than to suggest novel targets (which may still bear interest for basic research).

Personalized network target models could be probed first on retrospective datasets through basic research, cross-validations on retrospective data, or in virtual simulations (Meier et al., 2021). Once established, and only after careful validation, these concepts could inform prospective clinical trials investigating suitability of the proposed strategy to guide DBS programming and surgery. Given the reversible nature of DBS programming, the ethical implications would be lower, and this strategy could be applied before their use to refine surgical planning (which entails results that are not easily reversible). In DBS programming, new generations of directional electrode models could be leveraged to precisely steer the electrical field toward therapeutic and away from side-effect networks (Steigerwald et al., 2016). In view of possible dynamical changes in symptom constellations over time (e.g., following neurodegeneration), directional leads also allow for pre-implant planning with adjustments per time to anticipate future needs.

Ultimately, the proposed personalization concept could be translated into clinical care. In some cases, multi-focal strategies could help to modulate symptom-specific networks from different angles (Figee and Mayberg, 2021; Li et al., 2021, 2020; Scangos et al., 2021; Tyagi et al., 2019). The effectivity of a one-pass thalamic-subthalamic trajectory of DBS electrodes has, for instance, been investigated for concurrent treatment of tremor, rigidity and bradykinesia (Coenen et al., 2016; Neudorfer et al., 2019; Reinacher et al., 2018). Besides single-electrode strategies, co-stimulating distinct symptom networks could be achieved via multiple electrodes at non-adjacent sites, as during concomitant DBS to the GPi and STN (Sriram et al., 2014).

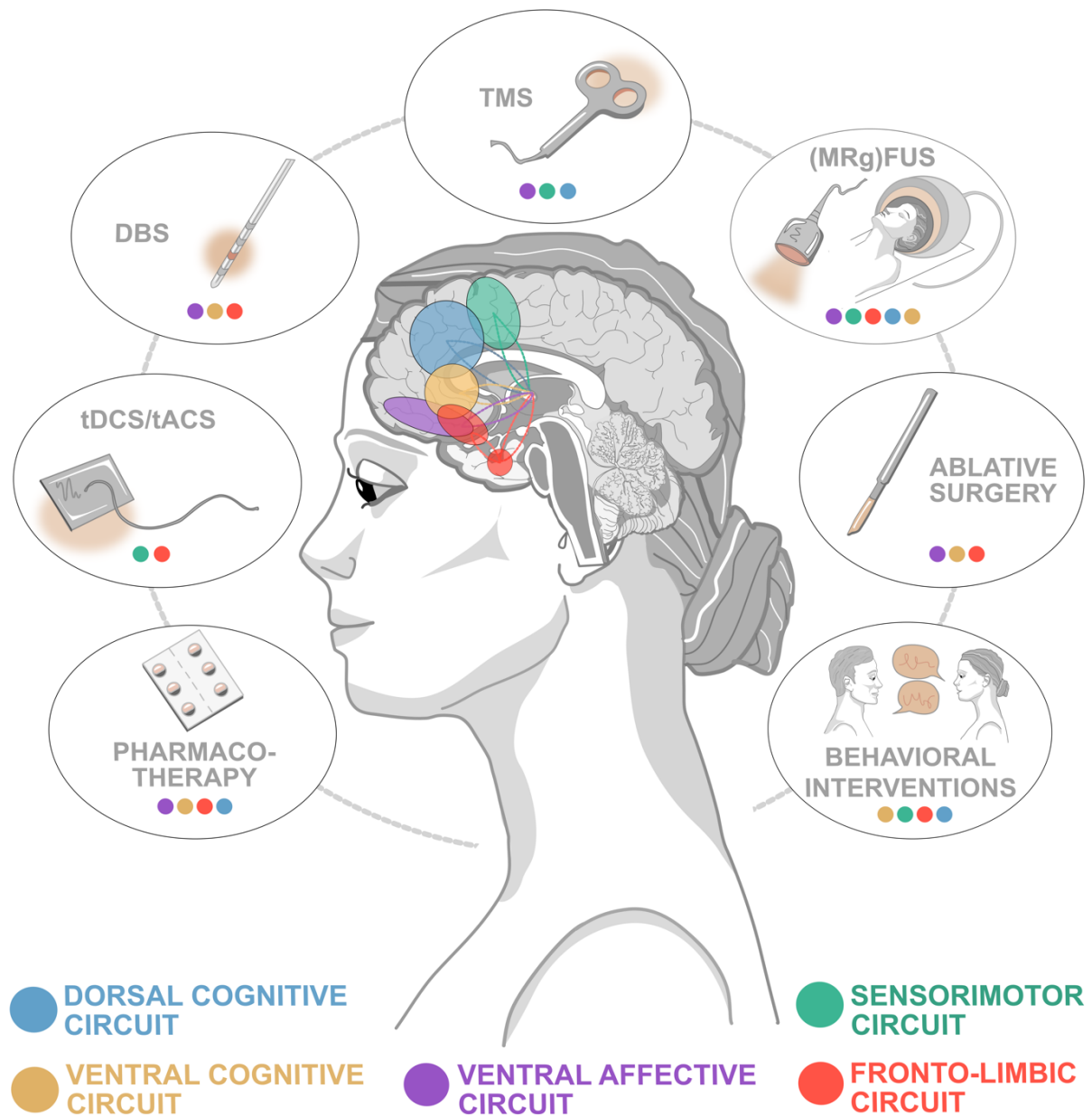

**Fig. 7:** Different treatment strategies (including neuromodulation modalities) may show differential effectiveness to impact specific symptom networks. This concept is visualized here on the example of obsessive-compulsive disorder (OCD) on the basis of an extensive literature integration proposed by Shephard et al. (2021). In OCD, these networks may be affected to varying degrees in individual patients – as a result forming heterogeneous OCD phenotypes. Once properly defined on a group level and re-identified in the individual patient, these may serve as potential targets to tailor treatment to a patient's unique symptom constellation. Crucially, specific (multimodal) treatment strategies may be particularly suited to treat each of these symptom networks, extending the proposed personalization framework beyond invasive neuromodulation. *Abbreviations:* DBS, deep brain stimulation; MRgFUS, Magnetic

Resonance-guided Focused Ultrasound; tACS, transcranial alternating current stimulation; tDCS, transcranial direct current stimulation; TMS, transcranial magnetic stimulation.

## 5. Limitations and future challenges

While further research is needed to carve out symptom-specific DBS network-effects, this blueprint constitutes an initial steppingstone toward circuit-based personalization of DBS. Many limitations apply which could, however, inspire future refinement of the suggested rationale and its underlying assumptions.

First, our approach assumes additive effects of DBS onto different symptom clusters and associated networks. By contrast, cortico-subcortical loops dynamically interact at different levels to generate the multifaceted psychological experiences and behaviors at the root of intricate phenotypes (Aoki et al., 2019; Guthrie et al., 2013; Kolomiets et al., 2001). On the other hand, such interactional effects may also prove advantageous when aiming to simultaneously account for multiple symptoms. Moreover, conclusions cannot be drawn about white-matter target engagement via the here-proposed methods. In future studies, potential biomarkers could be identified as a means of confirming post-operative propagation of electrical stimulation through target networks (Waters et al., 2018).

Second, current models of tissue activation are unable to account for the complex amalgamation of local and remote physiological DBS effects, comprising oscillatory, neuroprotective or neurochemical effects, ionisation of molecules, intracellular mechanisms, influence on glia cells, as well as synaptic plasticity or network reorganization (Ashkan et al., 2017; Herrington et al., 2016; Lozano et al., 2019; McIntyre et al., 2004; McIntyre and Anderson, 2016; Veerakumar and Berton, 2015a). Similarly, all DBS-related studies discussed applied high-frequency (i.e., >100 Hertz) stimulation, which produces effects resembling those of focal lesions (Neumann, 2021). However, this first-order approximation neglects the impact of frequency, pulse width, excitatory/inhibitory effects, or differential entrainment of fiber types of variable myelination. Effects of lower and/or variable and adaptive stimulation frequencies may have largely differing – or even opposing – effects (Benabid et al., 1991; Fox et al., 2014;

Krauss et al., 2021; Neudorfer et al., 2021). Further, different disorders – even those that respond to high-frequency DBS – may not necessarily present with a common pathology. Accordingly, methodological refinements of tissue activation modeling and pathophysiological investigations could increase predictive precision.

Third, collaboratively defining a library of DBS network targets challenges standardization and pathophysiological relevance in the design of symptom assessment batteries. Because symptom constructs are subjective, prone to fluctuations, and objective metrics or diagnostic criteria often lacking, quantifiability and relatability to connectomic biomarkers may be complicated (Odgers et al., 2009; Prange et al., 2019; Sullivan et al., 2021; Vergunst et al., 2013). A reliable and sparse collection of dimensional assessments will rely on critical examination and regular updating to avoid infinitesimal symptom axes with decreasing clinical utility (Morris and Cuthbert, 2012).

Fourth, symptom patterns and intensities may change over natural disease trajectories (Cilia et al., 2020; van Eeden et al., 2019). Also, postsurgical onset of stimulation effects may be delayed in some disorders (Krauss et al., 2004; Lozano et al., 2019; Veerakumar and Berton, 2015b), and tachyphylaxis, habituation (Benabid et al., 1996), or disease-modifying effects could emerge (Cif et al., 2013). Since many diseases treated by DBS technology are chronic or of neurodegenerative nature, preoperatively irrelevant symptoms might emerge and set the stage for future suboptimal personalization (Figee and Mayberg, 2021). While these factors underline a necessity to consider ongoing adjustments, the latter could also undermine the stability of DBS effects, favoring a "set-it-and-forget-it" strategy.

Fifth, neither will it be possible nor intended to treat each preoperative symptom with equal importance. Practical factors such as target/device certificates, psychosocial and socioeconomical circumstances, ethical concerns and the treatment's risk-benefit profile may critically weigh into interventional decisions (Accolla and Pollo, 2019; Kubu and Ford, 2017; Shephard et al., 2021). Also, neither will all patients or disease subtypes be eligible for DBS nor all symptoms accessible via this approach.

Last, precise DBS targeting is a critical first step to reduce outcome variance, which will promote the management of residual symptoms. Clearly, however, accounting for contributors to DBS outcome above and beyond electrode localization and associated connectivity will critically increase interventional precision.

## 6. Conclusions

The present whitepaper proposes a three-step framework for network-based personalization of neuromodulation based on i) establishing and ii) individualizing symptom-specific network targets. These can subsequently iii) be synthesized (or "*blended*") into personalized treatment targets tailored to the symptom profiles of an individual patient. While methodology for step i) has been developed and validated in the past, translational research is currently invested in refining methodological details for steps ii) and iii). In conclusion, we believe that this concept provides a powerful avenue toward advanced neuromodulation: First, it allows to systematically model *precise* connectomic predictors of stimulation effects onto various symptom dimensions. Second, it promotes interventional *personalization* to symptom heterogeneity.

## Acknowledgements

The authors would like to thank Prof. Dr. Volker A. Coenen for helpful comments and discussion on an initial draft of this manuscript.

## Funding

BH was supported via a scholarship from the Einstein Center for Neurosciences Berlin. SHS received support by the U.S. National Institute of Mental Health (K23MH121657), the Brain & Behavior Research Foundation, and the Baszucki Family Foundation. Further, this work received support by the Deutsche Forschungsgemeinschaft (DFG, German Research Foundation): grant numbers FI 2309/1-1 and FI 2309/2-1 to CF, grant number 347325977 to AAK, grant number 424778381–TRR 295 to AAK and AH, as well as Emmy Noether Stipend 410169619 to AH. HM was supported by NIH BRAIN UH3NS103550, R01MH102238, and the Hope For Depression Research Foundation. MDF was supported by the Nancy Lurie Marks Foundation, the Kaye Family Research Endowment, the Ellison/Baszucki Foundation, and the NIH (R01MH113929, R21MH126271, R56AG069086, R01MH115949, and R01AG060987). AH was supported by Deutsches Zentrum für Luft-und Raumfahrt (DynaSti grant within the EU Joint Programme Neurodegenerative Disease Research, JPND).

## Declaration of competing interests

SHS is a scientific consultant for Magnus Medical, provides clinical consulting for Kaizen Brain Center and Acacia Brain Center, has received research funding from Neuronetics Inc, and has received speaking fees from Brainsway Inc and Otsuka Pharmaceuticals. AAK reports lecture fees for Medtronic, Boston scientific, Abbott, Teva, Stada Pharm and Ipsen and serves in the advisory board for Medtronic. HM reports consulting and IP licensing fees from Abbott Labs. AH reports one-time lecture fees for Medtronic and Boston Scientific. BH, NR, CF, MDF and CN report no competing interests.

## CRediT author statement

BH and AH: Conceptualization, Writing – Original draft preparation, Visualization, Funding acquisition; NR: Writing – Reviewing and Editing, Visualization; SHS, CF, KB, AAK, HM, MDF, CN: Writing – Reviewing and Editing.

# References

Accolla, E.A., Pollo, C., 2019. Mood effects after deep brain stimulation for Parkinson's disease: An update. Front. Neurol. 10, 617. https://doi.org/10.3389/fneur.2019.00617

Akram, H., Dayal, V., Mahlknecht, P., Georgiev, D., Hyam, J., Foltynie, T., Limousin, P., De Vita, E., Jahanshahi, M., Ashburner, J., Behrens, T., Hariz, M., Zrinzo, L., 2018. Connectivity derived thalamic segmentation in deep brain stimulation for tremor. NeuroImage Clin. 18, 130–142. https://doi.org/10.1016/j.nicl.2018.01.008

Akram, H., Sotiropoulos, S.N., Jbabdi, S., Georgiev, D., Mahlknecht, P., Hyam, J., Foltynie, T., Limousin, P., De Vita, E., Jahanshahi, M., Hariz, M., Ashburner, J., Behrens, T., Zrinzo, L., 2017. Subthalamic deep brain stimulation sweet spots and hyperdirect cortical connectivity in Parkinson's disease. Neuroimage 158, 332–345. https://doi.org/10.1016/j.neuroimage.2017.07.012

Al-Fatly, B., Ewert, S., Kübler, D., Kroneberg, D., Horn, A., Kühn, A.A., 2019. Connectivity profile of thalamic deep brain stimulation to effectively treat essential tremor. Brain 142, 3086–3098. https://doi.org/10.1093/brain/awz236

Alexander, G., DeLong, M.R., Strick, P.L., 1986. Parallel organization of functionally segregated circuits linking basal ganglia and cortex. Annu. Rev. Neurosci. 9, 357–381. https://doi.org/10.1146/annurev.ne.09.030186.002041

Alexander, G.E., Crutcher, M.D., 1990. Functional architecture of basal ganglia circuits: neural substrates of parallel processing. Trends Neurosci. 13, 266–271. https://doi.org/10.1016/0166-2236(90)90107-l

Alho, E.J.L., Alho, A.T.D.L., Horn, A., Martin, M. da G.M., Edlow, B.L., Fischl, B., Nagy, J., Fonoff, E.T., Hamani, C., Heinsen, H., 2020. The ansa subthalamica: A neglected fiber tract. Mov. Disord. 35, 75–80. https://doi.org/10.1002/mds.27901

Alhourani, A., McDowell, M.M., Randazzo, M.J., Wozny, T.A., Kondylis, E.D., Lipski, W.J., Beck, S., Karp, J.F., Ghuman, A.S., Richardson, R.M., 2015. Network effects of deep brain stimulation. J. Neurophysiol. 114, 2105–2117.

https://doi.org/10.1152/jn.00275.2015

Allawala, A., Bijanki, K.R., Goodman, W., Cohn, J.F., Viswanathan, A., Yoshor, D., Borton, D.A., Pouratian, N., Sheth, S.A., 2021. A novel framework for network-targeted neuropsychiatric deep brain stimulation. Neurosurgery 89, 116–121. https://doi.org/10.1093/neuros/nyab112

Alonso, P., Cuadras, D., Gabriëls, L., Denys, D., Goodman, W., Greenberg, B.D., Jimenez-Ponce, F., Kuhn, J., Lenartz, D., Mallet, L., Nuttin, B., Real, E., Segalas, C., Schuurman, R., Du Montcel, S.T., Menchon, J.M., 2015. Deep brain stimulation for obsessive-compulsive disorder: A meta-analysis of treatment outcome and predictors of response. PLoS One 10, e0133591. https://doi.org/10.1371/journal.pone.0133591

Amunts, K., Lepage, C., Borgeat, L., Mohlberg, H., Dickscheid, T., Rousseau, M.-É., Bludau, S., Bazin, P.-L., Lewis, L.B., Oros-Peusquens, A.-M., Shah, N.J., Lippert, T., Zilles, K., Evans, A.C., 2013. BigBrain: An ultrahigh-resolution 3D human brain model. Science (80-. ). 340, 1472–1475. https://doi.org/10.1126/science.1235381

Anderson, J.S., Dhatt, H.S., Ferguson, M.A., Lopez-Larson, M., Schrock, L.E., House, P.A., Yurgelun-Todd, D., 2011. Functional connectivity targeting for deep brain stimulation in essential tremor. Am. J. Neuroradiol. 32, 1963–1968. https://doi.org/10.3174/ajnr.A2638

Aoki, S., Smith, J.B., Li, H., Yan, X., Igarashi, M., Coulon, P., Wickens, J.R., Ruigrok, T.J.H., Jin, X., 2019. An open cortico-basal ganglia loop allows limbic control over motor output via the nigrothalamic pathway. Elife 8, e49995. https://doi.org/10.7554/eLife.49995

Ashkan, K., Rogers, P., Bergman, H., Ughratdar, I., 2017. Insights into the mechanisms of deep brain stimulation. Nat. Rev. Neurol. 13, 548–554. https://doi.org/10.1038/nrneurol.2017.105

Baldermann, J.C., Hahn, L., Dembek, T.A., Kohl, S., Kuhn, J., Visser-Vandewalle, V., Horn, A., Huys, D., 2019a. Weight change after striatal/capsule deep brain stimulation relates to connectivity to the bed nucleus of the stria terminalis and hypothalamus. Brain Sci. 9, 264. https://doi.org/10.3390/brainsci9100264

Baldermann, J.C., Melzer, C., Zapf, A., Kohl, S., Timmermann, L., Tittgemeyer, M., Huys, D.,

Visser-Vandewalle, V., Kühn, A.A., Horn, A., Kuhn, J., 2019b. Connectivity profile predictive of effective deep brain stimulation in obsessive-compulsive disorder. Biol. Psychiatry 85, 735–743. https://doi.org/10.1016/j.biopsych.2018.12.019

Baldermann, J.C., Schüller, T., Kohl, S., Voon, V., Li, N., Hollunder, B., Figee, M., Haber, S.N., Sheth, S.A., Mosley, P.E., Huys, D., Johnson, K.A., Butson, C., Ackermans, L., van der Vlis, T.B., Leentjens, A.F.G., Barbe, M., Visser-Vandewalle, V., Kuhn, J., Horn, A., 2021. Connectomic deep brain stimulation for obsessive-compulsive disorder. Biol. Psychiatry. https://doi.org/10.1016/j.biopsych.2021.07.010

Barcia, J.A., Avecillas-Chasín, J.M., Nombela, C., Arza, R., García-Albea, J., Pineda-Pardo, J.A., Reneses, B., Strange, B.A., 2019. Personalized striatal targets for deep brain stimulation in obsessive-compulsive disorder. Brain Stimul. 12, 724–734. https://doi.org/10.1016/j.brs.2018.12.226

Bell, J., 2014. Stratified medicines: Towards better treatment for disease. Lancet 383, 3–5. https://doi.org/10.1016/S0140-6736(14)60115-X

Benabid, A.L., Pollak, P., Gao, D., Hoffmann, D., Limousin, P., Gay, E., Payen, I., Benazzouz, A., 1996. Chronic electrical stimulation of the ventralis intermedius nucleus of the thalamus as a treatment of movement disorders. J. Neurosurg. 84, 203–214. https://doi.org/10.3171/jns.1996.84.2.0203

Benabid, A.L., Pollak, P., Hoffmann, D., Gervason, C., Hommel, M., Perret, J.E., de Rougemont, J., Gao, D.M., 1991. Long-term suppression of tremor by chronic stimulation of the ventral intermediate thalamic nucleus. Lancet 337, 403–406. https://doi.org/10.1016/0140-6736(91)91175-T

Bergman, H., Wichmann, T., DeLong, M.R., 1990. Reversal of experimental parkinsonism by lesions of the subthalamic nucleus. Science (80-. ). 249, 1436–1438. https://doi.org/10.1126/science.2402638

Boutet, A., Germann, J., Gwun, D., Loh, A., Elias, G.J.B., Neudorfer, C., Paff, M., Horn, A., Kühn, A.A., Munhoz, R.P., Kalia, S.K., Hodaie, M., Kucharczyk, W., Fasano, A., Lozano, A.M., 2021. Sign-specific stimulation 'hot' and 'cold' spots in Parkinson's disease validated with machine learning. Brain Commun. 3, fcab027.
https://doi.org/10.1093/braincomms/fcab027

Braga, R.M., Buckner, R.L., 2017. Parallel interdigitated distributed networks within the individual estimated by intrinsic functional connectivity. Neuron 95, 457–471. https://doi.org/10.1016/j.neuron.2017.06.038

Calabrese, E., Hickey, P., Hulette, C., Zhang, J., Parente, B., Lad, S.P., Johnson, G.A., 2015. Postmortem diffusion MRI of the human brainstem and thalamus for deep brain stimulator electrode localization. Hum. Brain Mapp. 36, 3167–3178. https://doi.org/10.1002/hbm.22836

Calhoun, V.D., Kiehl, K.A., Pearlson, G.D., 2008. Modulation of temporally coherent brain networks estimated using ICA at rest and during cognitive tasks. Hum. Brain Mapp. 29, 828–838. https://doi.org/10.1002/hbm.20581

Casey, B.J., Craddock, N., Cuthbert, B.N., Hyman, S.E., Lee, F.S., Ressler, K.J., 2013. DSM-5 and RDoC: Progress in psychiatry research? Nat. Rev. Neurosci. 14, 810–814. https://doi.org/10.1038/nrn3621

Cash, R.F.H., Zalesky, A., Thomson, R.H., Tian, Y., Cocchi, L., Fitzgerald, P.B., 2019. Subgenual functional connectivity predicts antidepressant treatment response to transcranial magnetic stimulation: Independent validation and evaluation of personalization. Biol. Psychiatry 86, e5–e7. https://doi.org/10.1016/j.biopsych.2018.12.002

Cavallieri, F., Fraix, V., Bove, F., Mulas, D., Tondelli, M., Castrioto, A., Krack, P., Meoni, S., Schmitt, E., Lhommée, E., Bichon, A., Pélissier, P., Chevrier, E., Kistner, A., Seigneuret, E., Chabardès, S., Moro, E., 2021. Predictors of long-term outcome of subthalamic stimulation in Parkinson disease. Ann. Neurol. 89, 587–597. https://doi.org/10.1002/ana.25994

Cheng, W., Rolls, E.T., Qiu, J., Liu, W., Tang, Y., Huang, C.-C., Wang, X., Zhang, J., Lin, W., Zheng, L., Pu, J., Tsai, S.-J., Yang, A.C., Lin, C.-P., Wang, F., Xie, P., Feng, J., 2016. Medial reward and lateral non-reward orbitofrontal cortex circuits change in opposite directions in depression. Brain 139, 3296–3309. https://doi.org/10.1093/brain/aww255

Choi, K.S., Riva-Posse, P., Gross, R.E., Mayberg, H.S., 2015. Mapping the "depression switch" during intraoperative testing of subcallosal cingulate deep brain stimulation. JAMA Neurol. 72, 1252–1260. https://doi.org/10.1001/jamaneurol.2015.2564

Chudy, D., Deletis, V., Almahariq, F., Marčinković, P., Škrlin, J., Paradžik, V., 2018. Deep brain stimulation for the early treatment of the minimally conscious state and vegetative state: Experience in 14 patients. J. Neurosurg. 128, 1189–1198. https://doi.org/10.3171/2016.10.JNS161071

Cif, L., Ruge, D., Gonzalez, V., Limousin, P., Vasques, X., Hariz, M.I., Rothwell, J., Coubes, P., 2013. The influence of deep brain stimulation intensity and duration on symptoms evolution in an OFF stimulation dystonia study. Brain Stimul. 6, 500–505. https://doi.org/10.1016/j.brs.2012.09.005

Cilia, R., Cereda, E., Akpalu, A., Sarfo, F.S., Cham, M., Laryea, R., Obese, V., Oppon, K., Del Sorbo, F., Bonvegna, S., Zecchinelli, A.L., Pezzoli, G., 2020. Natural history of motor symptoms in Parkinson's disease and the long-duration response to levodopa. Brain 143, 2490–2501. https://doi.org/10.1093/brain/awaa181

Clementz, B.A., Sweeney, J.A., Hamm, J.P., Ivleva, E.I., Ethridge, L.E., Pearlson, G.D., Keshavan, M.S., Tamminga, C.A., 2016. Identification of distinct psychosis biotypes using brain-based biomarkers. Am. J. Psychiatry 173, 373–384. https://doi.org/10.1176/appi.ajp.2015.14091200

Coenen, V.A., Allert, N., Mädler, B., 2011. A role of diffusion tensor imaging fiber tracking in deep brain stimulation surgery: DBS of the dentato-rubro-thalamic tract (drt) for the treatment of therapy-refractory tremor. Acta Neurochir. (Wien). 153, 1579–1585. https://doi.org/10.1007/s00701-011-1036-z

Coenen, V.A., Honey, C.R., Hurwitz, T., Rahman, A.A., McMaster, J., Bürgel, U., Mädler, B., 2009. Medial forebrain bundle stimulation as a pathophysiological mechanism for hypomania in subthalamic nucleus deep brain stimulation for Parkinson's disease. Neurosurgery 64, 1106–1115. https://doi.org/10.1227/01.NEU.0000345631.54446.06

Coenen, V.A., Rijntjes, M., Prokop, T., Piroth, T., Amtage, F., Urbach, H., Reinacher, P.C., 2016. One-pass deep brain stimulation of dentato-rubro-thalamic tract and subthalamic nucleus for tremor-dominant or equivalent type Parkinson's disease. Acta Neurochir. (Wien). 158, 773–781. https://doi.org/10.1007/s00701-016-2725-4

Coenen, V.A., Sajonz, B., Prokop, T., Reisert, M., Piroth, T., Urbach, H., Jenkner, C., Reinacher, P.C., 2020. The dentato-rubro-thalamic tract as the potential common deep brain stimulation target for tremor of various origin: An observational case series. Acta Neurochir. (Wien). 162, 1053–1066. https://doi.org/10.1007/s00701-020-04248-2

Coenen, V.A., Schlaepfer, T.E., Goll, P., Reinacher, P.C., Voderholzer, U., Tebartz Van Elst, L., Urbach, H., Freyer, T., 2017. The medial forebrain bundle as a target for deep brain stimulation for obsessive-compulsive disorder. CNS Spectr. 22, 282–289. https://doi.org/10.1017/S1092852916000286

Coenen, V.A., Schumacher, L.V., Kaller, C., Schlaepfer, T.E., Reinacher, P.C., Egger, K., Urbach, H., Reisert, M., 2018. The anatomy of the human medial forebrain bundle: Ventral tegmental area connections to reward-associated subcortical and frontal lobe regions. NeuroImage Clin. 18, 770–783. https://doi.org/10.1016/j.nicl.2018.03.019

Corripio, I., Roldán, A., Sarró, S., McKenna, P.J., Alonso-Solís, A., Rabella, M., Díaz, A., Puigdemont, D., Pérez-Solà, V., Álvarez, E., Arévalo, A., Padilla, P.P., Ruiz-Idiago, J.M., Rodríguez, R., Molet, J., Pomarol-Clotet, E., Portella, M.J., 2020. Deep brain stimulation in treatment resistant schizophrenia: A pilot randomized cross-over clinical trial. EBioMedicine 51, 102568. https://doi.org/10.1016/j.ebiom.2019.11.029

Crocker, L.D., Heller, W., Warren, S.L., O'Hare, A.J., Infantolino, Z.P., Miller, G.A., 2013. Relationships among cognition, emotion, and motivation: Implications for intervention and neuroplasticity in psychopathology. Front. Hum. Neurosci. 7, 261. https://doi.org/10.3389/fnhum.2013.00261

Cury, R.G., Teixeira, M.J., Galhardoni, R., Silva, V., Iglesio, R., França, C., Arnaut, D., Fonoff, E.T., Barbosa, E.R., Ciampi de Andrade, D., 2020. Connectivity patterns of subthalamic stimulation influence pain outcomes in Parkinson's disease. Front. Neurol.

11, 9. https://doi.org/10.3389/fneur.2020.00009

Cuthbert, B.N., 2014. The RDoC framework: Facilitating transition from ICD/DSM to dimensional approaches that integrate neuroscience and psychopathology. World Psychiatry 13, 28–35. https://doi.org/10.1002/wps.20087

Cuthbert, B.N., Insel, T.R., 2013. Toward the future of psychiatric diagnosis: The seven pillars of RDoC. BMC Med. 11, 126. https://doi.org/10.1186/1741-7015-11-126

Darby, R.R., Horn, A., Cushman, F., Fox, M.D., 2018. Lesion network localization of criminal behavior. Proc. Natl. Acad. Sci. U. S. A. 115, 601–606. https://doi.org/10.1073/pnas.1706587115

De Almeida Marcelino, A.L., Horn, A., Krause, P., Kühn, A.A., Neumann, W.J., 2019. Subthalamic neuromodulation improves short-term motor learning in Parkinson's disease. Brain 142, 2198–2206. https://doi.org/10.1093/brain/awz152

Deeb, W., Rossi, P.J., Porta, M., Visser-Vandewalle, V., Servello, D., Silburn, P., Coyne, T., Leckman, J.F., Foltynie, T., Hariz, M., Joyce, E.M., Zrinzo, L., Kefalopoulou, Z., Welter, M.L., Karachi, C., Mallet, L., Houeto, J.-L., Shahed-Jimenez, J., Meng, F.-G., Klassen, B.T., Mogilner, A.Y., Pourfar, M.H., Kuhn, J., Ackermans, L., Kaido, T., Temel, Y., Gross, R.E., Walker, H.C., Lozano, A.M., Khandhar, S.M., Walter, B.L., Walter, E., Mari, Z., Changizi, B.K., Moro, E., Baldermann, J.C., Huys, D., Zauber, E.E., Schrock, L.E., Zhang, J.-G., Hu, W., Foote, K.D., Rizer, K., Mink, J.W., Woods, D.W., Gunduz, A., Okun, M.S., 2016. The international deep brain stimulation registry and database for Gilles de la Tourette syndrome: How does it work? Front. Neurosci. 10, 170. https://doi.org/10.3389/fnins.2016.00170

Deffains, M., Iskhakova, L., Katabi, S., Haber, S.N., Israel, Z., Bergman, H., 2016. Subthalamic, not striatal, activity correlates with basal ganglia downstream activity in normal and parkinsonian monkeys. Elife 5, e16443. https://doi.org/10.7554/eLife.16443

DeLong, M.R., 1990. Primate models of movement disorders of basal ganglia origin. Trends Neurosci. 13, 281–285. https://doi.org/10.1016/0166-2236(90)90110-V

Dinga, R., Schmaal, L., Penninx, B.W.J.H., van Tol, M.J., Veltman, D.J., van Velzen, L.,

Mennes, M., van der Wee, N.J.A., Marquand, A.F., 2019. Evaluating the evidence for biotypes of depression: Methodological replication and extension of Drysdale et al. (2017). NeuroImage Clin. 22, 101796. https://doi.org/10.1016/j.nicl.2019.101796

Dougherty, D.D., Rezai, A.R., Carpenter, L.L., Howland, R.H., Bhati, M.T., O'Reardon, J.P., Eskandar, E.N., Baltuch, G.H., Machado, A.D., Kondziolka, D., Cusin, C., Evans, K.C., Price, L.H., Jacobs, K., Pandya, M., Denko, T., Tyrka, A.R., Brelje, T., Deckersbach, T., Kubu, C., Malone, D.A.J., 2015. A randomized sham-controlled trial of deep brain stimulation of the ventral capsule/ventral striatum for chronic treatment-resistant depression. Biol. Psychiatry 78, 240–248. https://doi.org/10.1016/j.biopsych.2014.11.023

Drysdale, A.T., Grosenick, L., Downar, J., Dunlop, K., Mansouri, F., Meng, Y., Fetcho, R.N., Zebley, B., Oathes, D.J., Etkin, A., Schatzberg, A.F., Sudheimer, K., Keller, J., Mayberg, H.S., Gunning, F.M., Alexopoulos, G.S., Fox, M.D., Pascual-Leone, A., Voss, H.U., Casey, B.J., Dubin, M.J., Liston, C., 2017. Resting-state connectivity biomarkers define neurophysiological subtypes of depression. Nat. Med. 23, 28–38. https://doi.org/10.1038/nm.4246

Dunlop, B.W., Rajendra, J.K., Craighead, W.E., Kelley, M.E., McGrath, C.L., Choi, K.S., Kinkead, B., Nemeroff, C.B., Mayberg, H.S., 2017. Functional connectivity of the subcallosal cingulate cortex and differential outcomes to treatment with cognitive-behavioral therapy or antidepressant medication for major depressive disorder. Am. J. Psychiatry 174, 533–545. https://doi.org/10.1176/appi.ajp.2016.16050518

Eickhoff, S.B., Yeo, B.T.T., Genon, S., 2018. Imaging-based parcellations of the human brain. Nat. Rev. Neurosci. 19, 672–686. https://doi.org/10.1038/s41583-018-0071-7

Elias, G.J.B., Giacobbe, P., Boutet, A., Germann, J., Beyn, M.E., Gramer, R.M., Pancholi, A., Joel, S.E., Lozano, A.M., 2020. Probing the circuitry of panic with deep brain stimulation: Connectomic analysis and review of the literature. Brain Stimul. 13, 10–14. https://doi.org/10.1016/j.brs.2019.09.010

Etkin, A., 2019. Mapping causal circuitry in human depression. Biol. Psychiatry 86, 732–733.

https://doi.org/10.1016/j.biopsych.2019.09.009

Etkin, A., 2018. Addressing the causality gap in human psychiatric neuroscience. JAMA Psychiatry 75, 3–4. https://doi.org/10.1001/jamapsychiatry.2017.3610

Farrell, S.M., Green, A., Aziz, T., 2018. The current state of deep brain stimulation for chronic pain and its context in other forms of neuromodulation. Brain Sci. 8, 158. https://doi.org/10.3390/brainsci8080158

Fasano, A., Aquino, C.C., Krauss, J.K., Honey, C.R., Bloem, B.R., 2015. Axial disability and deep brain stimulation in patients with Parkinson disease. Nat. Rev. Neurol. 11, 98–110. https://doi.org/1038/nrneurol.2014.252 Introduction

Fearon, C., Lang, A.E., Espay, A.J., 2021. The logic and pitfalls of Parkinson's disease as "brain-first" versus "body-first" subtypes. Mov. Disord. 36, 594–598. https://doi.org/10.1002/mds.28493

Ferguson, M.A., Schaper, F.L.W.V.J., Cohen, A., Siddiqi, S., Merrill, S.M., Nielsen, J.A., Grafman, J., Urgesi, C., Fabbro, F., Fox, M.D., 2021. A neural circuit for spirituality and religiosity derived from patients with brain lesions. Biol. Psychiatry. https://doi.org/10.1016/j.biopsych.2021.06.016

Fernandes, B.S., Williams, L.M., Steiner, J., Leboyer, M., Carvalho, A.F., Berk, M., 2017. The new field of "precision psychiatry." BMC Med. 15, 80. https://doi.org/10.1186/s12916-017-0849-x

Figee, M., Mayberg, H., 2021. The future of personalized brain stimulation. Nat. Med. 27, 196–197. https://doi.org/10.1038/s41591-021-01243-7

Figee, M., Vink, M., De Geus, F., Vulink, N., Veltman, D.J., Westenberg, H., Denys, D., 2011. Dysfunctional reward circuitry in obsessive-compulsive disorder. Biol. Psychiatry 69, 867–874. https://doi.org/10.1016/j.biopsych.2010.12.003

Finn, E.S., Shen, X., Scheinost, D., Rosenberg, M.D., Huang, J., Chun, M.M., Papademetris, X., Constable, R.T., 2015. Functional connectome fingerprinting: Identifying individuals based on patterns of brain connectivity. Nat. Neurosci. 18, 1664–1671. https://doi.org/10.1038/nn.4135

Fox, M.D., Buckner, R.L., Liu, H., Mallar Chakravarty, M., Lozano, A.M., Pascual-Leone, A., 2014. Resting-state networks link invasive and noninvasive brain stimulation across diverse psychiatric and neurological diseases. Proc. Natl. Acad. Sci. U. S. A. 111, E4367–E4375. https://doi.org/10.1073/pnas.1405003111

Fox, M.D., Raichle, M.E., 2007. Spontaneous fluctuations in brain activity observed with functional magnetic resonance imaging. Nat. Rev. Neurosci. 8, 700–711. https://doi.org/10.1038/nrn2201

Frankemolle, A.M.M., Wu, J., Noecker, A.M., Voelcker-Rehage, C., Ho, J.C., Vitek, J.L., McIntyre, C.C., Alberts, J.L., 2010. Reversing cognitive-motor impairments in Parkinson's disease patients using a computational modelling approach to deep brain stimulation programming. Brain 133, 746–761. https://doi.org/10.1093/brain/awp315

Gardner, J., 2013. A history of deep brain stimulation: Technological innovation and the role of clinical assessment tools. Soc. Stud. Sci. 43, 707–728. https://doi.org/10.1177/0306312713483678

Gillan, C.M., Fineberg, N.A., Robbins, T.W., 2017. A trans-diagnostic perspective on obsessive-compulsive disorder. Psychol. Med. 47, 1528–1548. https://doi.org/10.1017/S0033291716002786

Gillan, C.M., Kosinski, M., Whelan, R., Phelps, E.A., Daw, N.D., 2016. Characterizing a psychiatric symptom dimension related to deficits in goal-directed control. Elife 5, e11305. https://doi.org/10.7554/eLife.11305

Gordon, E.M., Laumann, T.O., Adeyemo, B., Gilmore, A.W., Nelson, S.M., Dosenbach, N.U.F., Petersen, S.E., 2017a. Individual-specific features of brain systems identified with resting state functional correlations. Neuroimage 146, 918–939. https://doi.org/10.1016/j.neuroimage.2016.08.032

Gordon, E.M., Laumann, T.O., Adeyemo, B., Petersen, S.E., 2017b. Individual variability of the system-level organization of the human brain. Cereb. Cortex 27, 386–399. https://doi.org/10.1093/cercor/bhv239

Gordon, J.A., 2016. On being a circuit psychiatrist. Nat. Neurosci. 19, 1385–1386.

https://doi.org/10.1038/nn.4419

Görmezoğlu, M., Bouwens van der Vlis, T., Schruers, K., Ackermans, L., Polosan, M., Leentjens, A.F.G., 2020. Effectiveness, timing and procedural aspects of cognitive behavioral therapy after deep brain stimulation for therapy-resistant obsessive compulsive disorder: A systematic review. J. Clin. Med. 9, 2383. https://doi.org/10.3390/jcm9082383

Guthrie, M., Leblois, A., Garenne, A., Boraud, T., 2013. Interaction between cognitive and motor cortico-basal ganglia loops during decision making: A computational study. J. Neurophysiol. 109, 3025–3040. https://doi.org/10.1152/jn.00026.2013

Guzick, A., Hunt, P.J., Bijanki, K.R., Schneider, S.C., Sheth, S.A., Goodman, W.K., Storch, E.A., 2020. Improving long term patient outcomes from deep brain stimulation for treatment-refractory obsessive-compulsive disorder. Expert Rev. Neurother. 20, 95–107. https://doi.org/10.1080/14737175.2020.1694409

Harrison, B.J., Pujol, J., Cardoner, N., Deus, J., Alonso, P., López-Solà, M., Contreras-Rodríguez, O., Real, E., Segalàs, C., Blanco-Hinojo, L., Menchon, J.M., Soriano-Mas, C., 2013. Brain corticostriatal systems and the major clinical symptom dimensions of obsessive-compulsive disorder. Biol. Psychiatry 73, 321–328. https://doi.org/10.1016/j.biopsych.2012.10.006

Haynes, W.I.A., Haber, S.N., 2013. The organization of prefrontal-subthalamic inputs in primates provides an anatomical substrate for both functional specificity and integration: Implications for basal ganglia models and deep brain stimulation. J. Neurosci. 33, 4804–4814. https://doi.org/10.1523/JNEUROSCI.4674-12.2013

Herrington, T.M., Cheng, J.J., Eskandar, E.N., 2016. Mechanisms of deep brain stimulation. J. Neurophysiol. 115, 19–38. https://doi.org/10.1152/jn.00281.2015

Holtzheimer, P.E., Husain, M.M., Lisanby, S.H., Taylor, S.F., Whitworth, L.A., McClintock, S., Slavin, K. V., Berman, J., McKhann, G.M., Patil, P.G., Rittberg, B.R., Abosch, A., Pandurangi, A.K., Holloway, K.L., Lam, R.W., Honey, C.R., Neimat, J.S., Henderson, J.M., DeBattista, C., Rothschild, A.J., Pilitsis, J.G., Espinoza, R.T., Petrides, G.,

Mogilner, A.Y., Matthews, K., Peichel, D.L., Gross, R.E., Hamani, C., Lozano, A.M., Mayberg, H.S., 2017. Subcallosal cingulate deep brain stimulation for treatment-resistant depression: A multisite, randomised, sham-controlled trial. The Lancet Psychiatry 4, 839–849. https://doi.org/10.1016/S2215-0366(17)30371-1

Horn, A., 2019. The impact of modern-day neuroimaging on the field of deep brain stimulation. Curr. Opin. Neurol. 32, 511–520. https://doi.org/10.1097/WCO.0000000000000679

Horn, A., Fox, M.D., 2020. Opportunities of connectomic neuromodulation. Neuroimage 221, 117180. https://doi.org/10.1016/j.neuroimage.2020.117180

Horn, A., Reich, M., Vorwerk, J., Li, N., Wenzel, G., Fang, Q., Schmitz-Hübsch, T., Nickl, R., Kupsch, A., Volkmann, J., Kühn, A.A., Fox, M.D., 2017. Connectivity predicts deep brain stimulation outcome in Parkinson disease. Ann. Neurol. 82, 67–78. https://doi.org/10.1002/ana.24974

Horn, A., Wenzel, G., Irmen, F., Huebl, J., Li, N., Neumann, W.J., Krause, P., Bohner, G., Scheel, M., Kühn, A.A., 2019. Deep brain stimulation induced normalization of the human functional connectome in Parkinson's disease. Brain 142, 3129–3143. https://doi.org/10.1093/brain/awz239

Husain, M., 2017. Transdiagnostic neurology: Neuropsychiatric symptoms in neurodegenerative diseases. Brain 140, 1535–1536. https://doi.org/10.1093/brain/awx115

Insel, T., Cuthbert, B., Garvey, M., Heinssen, R., Pine, D., Quinn, K., Sanislow, C., Wang, P., 2010. Research Domain Criteria ( RDoC ): Toward a new classification framework for research on mental disorders. Am. J. Psychiatry 167, 748–751. https://doi.org/10.1176/appi.ajp.2010.09091379

Insel, T.R., 2014. The NIMH Research Domain Criteria (RDoC) Project: Precision medicine for psychiatry. Am. J. Psychiatry 171, 395–397. https://doi.org/10.1176/appi.ajp.2014.14020138

Irmen, F., Horn, A., Mosley, P., Perry, A., Petry-Schmelzer, J.N., Dafsari, H.S., Barbe, M.,

Visser-Vandewalle, V., Schneider, G.H., Li, N., Kübler, D., Wenzel, G., Kühn, A.A., 2020. Left prefrontal connectivity links subthalamic stimulation with depressive symptoms. Ann. Neurol. 87, 962–975. https://doi.org/10.1002/ana.25734

Ivleva, E.I., Clementz, B.A., Dutcher, A.M., Arnold, S.J.M., Jeon-Slaughter, H., Aslan, S., Witte, B., Poudyal, G., Lu, H., Meda, S.A., Pearlson, G.D., Sweeney, J.A., Keshavan, M.S., Tamminga, C.A., 2017. Brain structure biomarkers in the psychosis Bbotypes: Findings from the bipolar-schizophrenia network for intermediate phenotypes. Biol. Psychiatry 82, 26–39. https://doi.org/10.1016/j.biopsych.2016.08.030

Jakobs, M., Fomenko, A., Lozano, A.M., Kiening, K.L., 2019. Cellular, molecular, and clinical mechanisms of action of deep brain stimulation—a systematic review on established indications and outlook on future developments. EMBO Mol. Med. 11, e9575. https://doi.org/10.15252/emmm.201809575

Jeurissen, B., Descoteaux, M., Mori, S., Leemans, A., 2019. Diffusion MRI fiber tractography of the brain. NMR Biomed. 32, e3785. https://doi.org/10.1002/nbm.3785

Johnson, K.A., Duffley, G., Foltynie, T., Hariz, M., Zrinzo, L., Joyce, E.M., Akram, H., Servello, D., Galbiati, T.F., Bona, A., Porta, M., Meng, F.G., Leentjens, A.F.G., Gunduz, A., Hu, W., Foote, K.D., Okun, M.S., Butson, C.R., 2020. Basal ganglia pathways associated with therapeutic pallidal deep brain stimulation for Tourette syndrome. Biol. Psychiatry Cogn. Neurosci. Neuroimaging. https://doi.org/10.1016/j.bpsc.2020.11.005

Kapur, S., Phillips, A.G., Insel, T.R., 2012. Why has it taken so long for biological psychiatry to develop clinical tests and what to do about it? Mol. Psychiatry 17, 1174–1179. https://doi.org/10.1038/mp.2012.105

Kelley, M.E., Choi, K.S., Rajendra, J.K., Craighead, W.E., Rakofsky, J.J., Dunlop, B.W., Mayberg, H.S., 2021. Establishing evidence for clinical utility of a neuroimaging biomarker in major depressive disorder: Prospective testing and implementation challenges. Biol. Psychiatry 90, 236–242. https://doi.org/10.1016/j.biopsych.2021.02.966

Kolomiets, B.P., Deniau, J.M., Mailly, P., Ménétrey, A., Glowinski, J., Thierry, A.M., 2001.

Segregation and convergence of information flow through the cortico-subthalamic pathways. J. Neurosci. 21, 5764–5772. https://doi.org/10.1523/jneurosci.21-15-05764.2001

Kong, R., Li, J., Orban, C., Sabuncu, M.R., Liu, H., Schaefer, A., Sun, N., Zuo, X.-N., Holmes, A.J., Eickhoff, S.B., Yeo, B.T.T., 2019. Spatial topography of individual-specific cortical networks predicts human cognition, personality, and emotion. Cereb. Cortex 29, 2533–2551. https://doi.org/10.1093/cercor/bhy123

Korgaonkar, M.S., Rekshan, W., Gordon, E., Rush, A.J., Williams, L.M., Blasey, C., Grieve, S.M., 2015. Magnetic resonance imaging measures of brain structure to predict antidepressant treatment outcome in major depressive disorder. EBioMedicine 2, 37–45. https://doi.org/10.1016/j.ebiom.2014.12.002

Krauss, J.K., Lipsman, N., Aziz, T., Boutet, A., Brown, P., Chang, J.W., Davidson, B., Grill, W.M., Hariz, M.I., Horn, A., Schulder, M., Mammis, A., Tass, P.A., Volkmann, J., Lozano, A.M., 2021. Technology of deep brain stimulation: current status and future directions. Nat. Rev. Neurol. 17, 75–87. https://doi.org/10.1038/s41582-020-00426-z

Krauss, J.K., Yianni, J., Loher, T.J., Aziz, T.Z., 2004. Deep brain stimulation for dystonia. J. Clin. Neurophysiol. 21, 18–30. https://doi.org/10.1097/00004691-200401000-00004

Kubu, C.S., Ford, P.J., 2017. Clinical ethics in the context of deep brain stimulation for movement disorders. Arch. Clin. Neuropsychol. 32, 829–839. https://doi.org/10.1093/arclin/acx088

Kuhn, J., Hardenacke, K., Lenartz, D., Gruendler, T., Ullsperger, M., Bartsch, C., Mai, J.K., Zilles, K., Bauer, A., Matusch, A., Schulz, R.-J., Noreik, M., Bührle, C.P., Maintz, D., Woopen, C., Häussermann, P., Hellmich, M., Klosterkötter, J., Wiltfang, J., Maarouf, M., Freund, H.-J., Sturm, V., 2015. Deep brain stimulation of the nucleus basalis of Meynert in Alzheimer's dementia. Mol. Psychiatry 20, 353–360. https://doi.org/10.1038/mp.2014.32

Lansdall, C.J., Coyle-Gilchrist, I.T.S., Jones, P.S., Vázquez Rodríguez, P., Wilcox, A., Wehmann, E., Dick, K.M., Robbins, T.W., Rowe, J.B., 2017. Apathy and impulsivity in frontotemporal lobar degeneration syndromes. Brain 140, 1792–1807. https://doi.org/10.1093/brain/awx101

Laxton, A.W., Tang-Wai, D.F., McAndrews, M.P., Zumsteg, D., Wennberg, R., Keren, R., Wherrett, J., Naglie, G., Hamani, C., Smith, G.S., Lozano, A.M., 2010. A phase I trial of deep brain stimulation of memory circuits in Alzheimer's disease. Ann. Neurol. 68, 521–534. https://doi.org/10.1002/ana.22089

Lee, D.J., Lozano, C.S., Dallapiazza, R.F., Lozano, A.M., 2019. Current and future directions of deep brain stimulation for neurological and psychiatric disorders. J. Neurosurg. 131, 333–342. https://doi.org/10.3171/2019.4.JNS181761

Li, N., Baldermann, J.C., Kibleur, A., Treu, S., Akram, H., Elias, G.J.B., Boutet, A., Lozano, A.M., Al-Fatly, B., Strange, B., Barcia, J.A., Zrinzo, L., Joyce, E., Chabardes, S., Visser-Vandewalle, V., Polosan, M., Kuhn, J., Kühn, A.A., Horn, A., 2020. A unified connectomic target for deep brain stimulation in obsessive-compulsive disorder. Nat. Commun. 11, 3364. https://doi.org/10.1038/s41467-020-16734-3

Li, N., Hollunder, B., Baldermann, J.C., Kibleur, A., Treu, S., Akram, H., Al-Fatly, B., Strange, B.A., Barcia, J.A., Zrinzo, L., Joyce, E.M., Chabardes, S., Visser-Vandewalle, V., Polosan, M., Kuhn, J., Kühn, A.A., Horn, A., 2021. A unified functional network target for deep brain stimulation in obsessive-compulsive disorder. Biol. Psychiatry. https://doi.org/10.1016/j.biopsych.2021.04.006

Liang, S., Deng, W., Li, X., Greenshaw, A.J., Wang, Q., Li, M., Ma, X., Bai, T.-J., Bo, Q.-J., Cao, J., Chen, G.-M., Chen, W., Cheng, C., Cheng, Y.-Q., Cui, X.-L., Duan, J., Fang, Y.-R., Gong, Q.-Y., Guo, W.-B., Hou, Z.-H., Hu, L., Kuang, L., Li, F., Li, K.-M., Liu, Y.-S., Liu, Z.-N., Long, Y.-C., Luo, Q.-H., Meng, H.-Q., Peng, D.-H., Qiu, H.-A., Qiu, J., Shen, Y.-D., Shi, Y.-S., Si, T.-M., Wang, C.-Y., Wang, F., Wang, K., Wang, L., Wang, X., Wang, Y., Wu, X.-P., Wu, X.-R., Xie, C.-M., Xie, G.-R., Xie, H.-Y., Xie, P., Xu, X.-F., Yang, H., Yang, J., Yu, H., Yao, J.-S., Yao, S.-Q., Yin, Y.-Y., Yuan, Y.-G., Zang, Y.-F., Zhang, A.-X., Zhang, H., Zhang, K.-R., Zhang, Z.-J., Zhao, J.-P., Zhou, R.-B., Zhou, Y.-T., Zou, C.-J., Zuo, X.-N., Yan, C.-G., Li, T., 2020. Biotypes of major depressive disorder: Neuroimaging evidence from resting-state default mode network patterns, NeuroImage: Clinical. The Author(s). https://doi.org/10.1016/j.nicl.2020.102514

Limousin, P., Foltynie, T., 2019. Long-term outcomes of deep brain stimulation in Parkinson disease. Nat. Rev. Neurol. 15, 234–242. https://doi.org/10.1038/s41582-019-0145-9

Llera, A., Wolfers, T., Mulders, P., Beckmann, C.F., 2019. Inter-individual differences in human brain structure and morphology link to variation in demographics and behavior. Elife 8, e44443. https://doi.org/10.7554/eLife.44443

Lofredi, R., Auernig, G.C., Irmen, F., Nieweler, J., Neumann, W.J., Horn, A., Schneider, G.H., Kühn, A.A., 2021. Subthalamic stimulation impairs stopping of ongoing movements. Brain 144, 44–52. https://doi.org/10.1093/brain/awaa341

Lozano, A.M., Lipsman, N., 2013. Probing and regulating dysfunctional circuits using deep brain stimulation. Neuron 77, 406–424. https://doi.org/10.1016/j.neuron.2013.01.020

Lozano, A.M., Lipsman, N., Bergman, H., Brown, P., Chabardes, S., Chang, J.W., Matthews, K., McIntyre, C.C., Schlaepfer, T.E., Schulder, M., Temel, Y., Volkmann, J., Krauss, J.K., 2019. Deep brain stimulation: Current challenges and future directions. Nat. Rev. Neurol. 15, 148–160. https://doi.org/10.1038/s41582-018-0128-2

Maier-Hein, K.H., Neher, P.F., Houde, J.C., Côté, M.A., Garyfallidis, E., Zhong, J., Chamberland, M., Yeh, F.C., Lin, Y.C., Ji, Q., Reddick, W.E., Glass, J.O., Chen, D.Q., Feng, Y., Gao, C., Wu, Y., Ma, J., Renjie, H., Li, Q., Westin, C.F., Deslauriers-Gauthier, S., González, J.O.O., Paquette, M., St-Jean, S., Girard, G., Rheault, F., Sidhu, J., Tax, C.M.W., Guo, F., Mesri, H.Y., Dávid, S., Froeling, M., Heemskerk, A.M., Leemans, A., Boré, A., Pinsard, B., Bedetti, C., Desrosiers, M., Brambati, S., Doyon, J., Sarica, A., Vasta, R., Cerasa, A., Quattrone, A., Yeatman, J., Khan, A.R., Hodges, W., Alexander, S., Romascano, D., Barakovic, M., Auría, A., Esteban, O., Lemkaddem, A., Thiran, J.P., Cetingul, H.E., Odry, B.L., Mailhe, B., Nadar, M.S., Pizzagalli, F., Prasad, G., Villalon-Reina, J.E., Galvis, J., Thompson, P.M., Requejo, F.D.S., Laguna, P.L., Lacerda, L.M., Barrett, R., Dell'Acqua, F., Catani, M., Petit, L., Caruyer, E., Daducci, A., Dyrby, T.B., Holland-Letz, T., Hilgetag, C.C., Stieltjes, B., Descoteaux, M., 2017. The challenge of mapping the human connectome based on diffusion tractography. Nat. Commun. 8, 1349. https://doi.org/10.1038/s41467-017-01285-x

Mantione, M., Nieman, D.H., Figee, M., Denys, D., 2014. Cognitive-behavioural therapy augments the effects of deep brain stimulation in obsessive-compulsive disorder. Psychol. Med. 44, 3515–3522. https://doi.org/10.1017/S0033291714000956

Marquand, A.F., Rezek, I., Buitelaar, J., Beckmann, C.F., 2016. Understanding heterogeneity in clinical cohorts using normative models: Beyond case-control studies. Biol. Psychiatry 80, 552–561. https://doi.org/10.1016/j.biopsych.2015.12.023

Mataix-Cols, D., Wooderson, S., Lawrence, N., Brammer, M.J., Speckens, A., Phillips, M.L., 2004. Distinct neural correlates of washing, checking, and hoarding symptom dimensions in obsessive-compulsive disorder. Arch. Gen. Psychiatry 61, 564–576. https://doi.org/10.1001/archpsyc.61.6.564

Mayberg, H.S., Lozano, A.M., Voon, V., McNeely, H.E., Seminowicz, D., Hamani, C., Schwalb, J.M., Kennedy, S.H., 2005. Deep brain stimulation for treatment-resistant depression. Neuron 45, 651–660. https://doi.org/10.1016/j.neuron.2005.02.014

McGrath, C.L., Kelley, M.E., Holtzheimer, P.E., Dunlop, B.W., Craighead, W.E., Franco, A.R., Craddock, R.C., Mayberg, H.S., 2013. Toward a neuroimaging treatment selection biomarker for major depressive disorder. JAMA Psychiatry 70, 821–829. https://doi.org/10.1001/jamapsychiatry.2013.143

McIntyre, C.C., Anderson, R.W., 2016. Deep brain stimulation mechanisms: The control of network activity via neurochemistry modulation. J. Neurochem. 139, 338–345. https://doi.org/10.1111/jnc.13649

McIntyre, C.C., Grill, W.M., Sherman, D.L., Thakor, N. V., 2004. Cellular effects of deep brain stimulation: Model-based analysis of activation and inhibition. J. Neurophysiol. 91, 1457–1469. https://doi.org/10.1152/jn.00989.2003

Meier, J.M., Perdikis, D., Blickensdörfer, A., Stefanovski, L., Liu, Q., Maith, O., Dinkelbach, H.Ü., Baladron, J., Hamker, F.H., Ritter, P., 2021. Virtual deep brain stimulation: Multiscale co-simulation of a spiking basal ganglia model and a whole-brain mean-field model with The Virtual Brain. bioRxiv.

Merikangas, K.R., Calkins, M.E., Burstein, M., He, J.-P., Chiavacci, R., Lateef, T., Ruparel, K., Gur, R.C., Lehner, T., Hakonarson, H., Gur, R.E., 2015. Comorbidity of physical and mental disorders in the neurodevelopmental genomics cohort study. Pediatrics 135, e927–e938. https://doi.org/10.1542/peds.2014-1444

Mestre, T.A., Fereshtehnejad, S.-M., Berg, D., Bohnen, N.I., Dujardin, K., Erro, R., Espay, A.J., Halliday, G., Van Hilten, J.J., Hu, M.T., Jeon, B., Klein, C., Leentjens, A.F.G., Marinus, J., Mollenhauer, B., Postuma, R., Rajalingam, R., Rodríguez-Violante, M., Simuni, T., Surmeier, D.J., Weintraub, D., McDermott, M.P., Lawton, M., Marras, C., 2021. Parkinson's disease subtypes: Critical appraisal and recommendations. J. Parkinsons. Dis. 11, 395–404. https://doi.org/10.3233/JPD-202472

Middlebrooks, E.H., Grewal, S.S., Stead, M., Lundstrom, B.N., Worrell, G.A., Van Gompel, J.J., 2018. Differences in functional connectivity profiles as a predictor of response to anterior thalamic nucleus deep brain stimulation for epilepsy: A hypothesis for the mechanism of action and a potential biomarker for outcomes. Neurosurg. Focus 45, E7. https://doi.org/10.3171/2018.5.FOCUS18151

Miterko, L.N., Lin, T., Zhou, J., van der Heijden, M.E., Beckinghausen, J., White, J.J., Sillitoe, R. V., 2021. Neuromodulation of the cerebellum rescues movement in a mouse model of ataxia. Nat. Commun. 12, 1295. https://doi.org/10.1038/s41467-021-21417-8

Moro, E., LeReun, C., Krauss, J.K., Albanese, A., Lin, J.-P., Walleser Autiero, S., Brionne, T.C., Vidailhet, M., 2017. Efficacy of pallidal stimulation in isolated dystonia: A systematic review and meta-analysis. Eur. J. Neurol. 24, 552–560. https://doi.org/10.1111/ene.13255

Morris, S.E., Cuthbert, B.N., 2012. Research domain criteria: Cognitive systems, neural circuits, and dimensions of behavior. Dialogues Clin. Neurosci. 14, 29–37. https://doi.org/10.31887/dcns.2012.14.1/smorris

Mosley, P.E., Paliwal, S., Robinson, K., Coyne, T., Silburn, P., Tittgemeyer, M., Stephan, K.E., Perry, A., Breakspear, M., 2020. The structural connectivity of subthalamic deep brain stimulation correlates with impulsivity in Parkinson's disease. Brain 143, 2235–2254. https://doi.org/10.1093/brain/awaa148

Mosley, P.E., Windels, F., Morris, J., Coyne, T., Marsh, R., Giorni, A., Mohan, A., Sachdev, P., O'Leary, E., Boschen, M., Sah, P., Silburn, P.A., 2021. A randomised, double-blind, sham-controlled trial of deep brain stimulation of the bed nucleus of the stria terminalis for treatment-resistant obsessive-compulsive disorder. Transl. Psychiatry 11, 190. https://doi.org/10.1038/s41398-021-01307-9

Neudorfer, C., Chow, C.T., Boutet, A., Loh, A., Germann, J., Elias, G.J., Hutchison, W.D., Lozano, A.M., 2021. Kilohertz-frequency stimulation of the nervous system: A review of underlying mechanisms. Brain Stimul. 14, 513–530. https://doi.org/10.1016/j.brs.2021.03.008

Neudorfer, C., Hinzke, M., Hunsche, S., El Majdoub, F., Lozano, A., Maarouf, M., 2019. Combined deep brain stimulation of subthalamic nucleus and ventral intermediate thalamic nucleus in tremor-dominant Parkinson's disease using a parietal approach. Neuromodulation 22, 493–502. https://doi.org/10.1111/ner.12943

Neumann, W.J., 2021. Neurophysiological mechanisms of DBS from a connectomic perspective, in: Horn, A. (Ed.), Connectomic Deep Brain Stimulation. Elsevier Academic Press, Cambridge, MA.

Neumann, W.J., Schroll, H., De Almeida Marcelino, A.L., Horn, A., Ewert, S., Irmen, F., Krause, P., Schneider, G.-H., Hamker, F., Kühn, A.A., 2018. Functional segregation of basal ganglia pathways in Parkinson's disease. Brain 141, 2655–2669. https://doi.org/10.1093/brain/awy206

Nusslock, R., Alloy, L.B., 2017. Reward processing and mood-related symptoms: An RDoC and translational neuroscience perspective. J. Affect. Disord. 216, 3–16. https://doi.org/10.1016/j.jad.2017.02.001

Odgers, C.L., Mulvey, E.P., Skeem, J.L., Gardner, W., Lidz, C.W., Schubert, C., 2009. Capturing the ebb and flow of psychiatric symptoms with dynamical systems models. Am. J. Psychiatry 166, 575–582. https://doi.org/10.1176/appi.ajp.2008.08091398

Okromelidze, L., Tsuboi, T., Eisinger, R.S., Burns, M.R., Charbel, M., Rana, M., Grewal, S.S., Lu, C.-Q., Almeida, L., Foote, K.D., Okun, M.S., Middlebrooks, E.H., 2020. Functional and structural connectivity patterns associated with clinical outcomes in deep brain stimulation of the globus pallidus internus for generalized dystonia. Am. J. Neuroradiol. 41, 508–514. https://doi.org/10.3174/ajnr.A6429

Okun, M.S., Tagliati, M., Pourfar, M., Fernandez, H.H., Rodriguez, R.L., Alterman, R.L., Foote, K.D., 2005. Management of referred deep brain stimulation failures. Arch. Neurol. 62, 1250–1255. https://doi.org/10.1001/archneur.62.8.noc40425

Pauls, K.A.M., Krauss, J.K., Kämpfer, C.E., Kühn, A.A., Schrader, C., Südmeyer, M., Allert, N., Benecke, R., Blahak, C., Boller, J.K., Fink, G.R., Fogel, W., Liebig, T., El Majdoub, F., Mahlknecht, P., Kessler, J., Mueller, J., Voges, J., Wittstock, M., Wolters, A., Maarouf, M., Moro, E., Volkmann, J., Bhatia, K.P., Timmermann, L., 2017. Causes of failure of pallidal deep brain stimulation in cases with pre-operative diagnosis of isolated dystonia. Park. Relat. Disord. 43, 38–48. https://doi.org/10.1016/j.parkreldis.2017.06.023

Perlis, R.H., 2011. Translating biomarkers to clinical practice. Mol. Psychiatry 16, 1076–1087. https://doi.org/10.1038/mp.2011.63

Petersen, M. V, Mlakar, J., Haber, S.N., Parent, M., Smith, Y., Strick, P.L., Griswold, M.A., McIntyre, C.C., 2019. Holographic reconstruction of axonal pathways in the human brain. Neuron 104, 1056-1064.e3. https://doi.org/10.1016/j.neuron.2019.09.030.

Pilitsis, J.G., Metman, L.V., Toleikis, J.R., Hughes, L.E., Sani, S.B., Bakay, R.A.E., 2008. Factors involved in long-term efficacy of deep brain stimulation of the thalamus for essential tremor. J. Neurosurg. 109, 640–646. https://doi.org/10.3171/JNS/2008/109/10/0640

Plana-Ripoll, O., Pedersen, C.B., Holtz, Y., Benros, M.E., Dalsgaard, S., De Jonge, P., Fan, C.C., Degenhardt, L., Ganna, A., Greve, A.N., Gunn, J., Iburg, K.M., Kessing, L.V., Lee, B.K., Lim, C.C.W., Mors, O., Nordentoft, M., Prior, A., Roest, A.M., Saha, S., Schork, A., Scott, J.G., Scott, K.M., Stedman, T., Sørensen, H.J., Werge, T., Whiteford, H.A.,

Laursen, T.M., Agerbo, E., Kessler, R.C., Mortensen, P.B., McGrath, J.J., 2019. Exploring comorbidity within mental disorders among a danish national population. JAMA Psychiatry 76, 259–270. https://doi.org/10.1001/jamapsychiatry.2018.3658

Prange, S., Danaila, T., Laurencin, C., Caire, C., Metereau, E., Merle, H., Broussolle, E., Maucort-Boulch, D., Thobois, S., 2019. Age and time course of long-term motor and nonmotor complications in Parkinson disease. Neurology 92, e148–e160. https://doi.org/10.1212/WNL.0000000000006737

Price, R.B., Lane, S., Gates, K., Kraynak, T.E., Horner, M.S., Thase, M.E., Siegle, G.J., 2017. Parsing heterogeneity in the brain connectivity of depressed and healthy adults during positive mood. Biol. Psychiatry 81, 347–357. https://doi.org/10.1016/j.biopsych.2016.06.023

Priori, A., 2015. Technology for deep brain stimulation at a gallop. Mov. Disord. 30, 1206–1212. https://doi.org/10.1002/mds.26253

Reinacher, P.C., Amtage, F., Rijntjes, M., Piroth, T., Prokop, T., Jenkner, C., Kätzler, J., Coenen, V.A., 2018. One Pass Thalamic and Subthalamic Stimulation for Patients with Tremor-Dominant Idiopathic Parkinson Syndrome (OPINION): Protocol for a Randomized, Active-Controlled, Double-Blinded Pilot Trial. JMIR Res. Protoc. 7, e36. https://doi.org/10.2196/resprot.8341

Reisert, M., Mader, I., Anastasopoulos, C., Weigel, M., Schnell, S., Kiselev, V., 2011. Global fiber reconstruction becomes practical. Neuroimage 54, 955–962. https://doi.org/10.1016/j.neuroimage.2010.09.016

Riva-Posse, P., Choi, K.S., Holtzheimer, P.E., Crowell, A.L., Garlow, S.J., Rajendra, J.K., McIntyre, C.C., Gross, R.E., Mayberg, H.S., 2018. A connectomic approach for subcallosal cingulate deep brain stimulation surgery: Prospective targeting in treatment-resistant depression. Mol. Psychiatry 23, 843–849. https://doi.org/10.1038/mp.2017.59

Riva-Posse, P., Choi, K.S., Holtzheimer, P.E., McIntyre, C.C., Gross, R.E., Chaturvedi, A., Crowell, A.L., Garlow, S.J., Rajendra, J.K., Mayberg, H.S., 2014. Defining critical white matter pathways mediating successful subcallosal cingulate deep brain stimulation for treatment-resistant depression. Biol. Psychiatry 76, 963–969.

https://doi.org/10.1016/j.biopsych.2014.03.029

Robbins, T.W., Vaghi, M.M., Banca, P., 2019. Obsessive-compulsive disorder: Puzzles and prospects. Neuron 102, 27–47. https://doi.org/10.1016/j.neuron.2019.01.046

Rodriguez-Oroz, M.C., Moro, E., Krack, P., 2012. Long-term outcomes of surgical therapies for Parkinson's disease. Mov. Disord. 27, 1718–1728. https://doi.org/10.1002/mds.25214

Rohlfing, T., Kroenke, C.D., Sullivan, E. V., Dubach, M.F., Bowden, D.M., Grant, K.A., Pfefferbaum, A., 2012. The INIA19 template and NeuroMaps atlas for primate brain image parcellation and spatial normalization. Front. Neuroinform. 6, 27. https://doi.org/10.3389/fninf.2012.00027

Scangos, K.W., Makhoul, G.S., Sugrue, L.P., Chang, E.F., Krystal, A.D., 2021. State-dependent responses to intracranial brain stimulation in a patient with depression. Nat. Med. 27, 229–231. https://doi.org/10.1038/s41591-020-01175-8

Schlaepfer, T.E., Bewernick, B.H., Kayser, S., Hurlemann, R., Coenen, V.A., 2014. Deep brain stimulation of the human reward system for major depression - Rationale, outcomes and outlook. Neuropsychopharmacology 39, 1303–1314. https://doi.org/10.1038/npp.2014.28

Schlaepfer, T.E., Bewernick, B.H., Kayser, S., Mädler, B., Coenen, V.A., 2013. Rapid effects of deep brain stimulation for treatment-resistant major depression. Biol. Psychiatry 73, 1204–1212. https://doi.org/10.1016/j.biopsych.2013.01.034

Shalash, A., Alexoudi, A., Knudsen, K., Volkmann, J., Mehdorn, M., Deuschl, G., 2014. The impact of age and disease duration on the long term outcome of neurostimulation of the subthalamic nucleus. Park. Relat. Disord. 20, 47–52. https://doi.org/10.1016/j.parkreldis.2013.09.014

Sharma, A., Szeto, K., Desilets, A.R., 2012. Efficacy and safety of deep brain stimulation as an adjunct to pharmacotherapy for the treatment of Parkinson disease. Ann. Pharmacother. 46, 248–254. https://doi.org/10.1345/aph.1Q508

Shephard, E., Stern, E.R., van den Heuvel, O.A., Costa, D.L.C., Batistuzzo, M.C., Godoy, P.B.G., Lopes, A.C., Brunoni, A.R., Hoexter, M.Q., Shavitt, R.G., Reddy, Y.C.J., Lochner, C., Stein, D.J., Simpson, H.B., Miguel, E.C., 2021. Toward a neurocircuit-based taxonomy to guide treatment of obsessive–compulsive disorder. Mol. Psychiatry. https://doi.org/10.1038/s41380-020-01007-8

Sheth, S.A., Neal, J., Tangherlini, F., Mian, M.K., Gentil, A., Cosgrove, G.R., Eskandar, E.N., Dougherty, D.D., 2013. Limbic system surgery for treatment-refractory obsessive-compulsive disorder: A prospective long-term follow-up of 64 patients. J. Neurosurg. 118, 491–497. https://doi.org/10.3171/2012.11.JNS12389

Siddiqi, S.H., Schaper, F.L.W.V.J., Horn, A., Hsu, J., Padmanabhan, J.L., Brodtmann, A., Cash, R.F.H., Corbetta, M., Choi, K.S., Dougherty, D.D., Egorova, N., Fitzgerald, P.B., George, M.S., Gozzi, S.A., Irmen, F., Kühn, A.A., Johnson, K.A., Naidech, A.M., Pascual-Leone, A., Phan, T.G., Rouhl, R.P.W., Taylor, S.F., Voss, J.L., Zalesky, A., Grafmann, J.H., Mayberg, H.S., Fox, M.D., 2021. Brain stimulation and brain lesions converge on common causal circuits in neuropsychiatric disease. Nat. Hum. Behav. https://doi.org/10.1038/s41562-021-01161-1

Siddiqi, S.H., Taylor, S.F., Cooke, D., Pascual-Leone, A., George, M.S., Fox, M.D., 2020. Distinct symptom-specific treatment targets for circuit-based neuromodulation. Am. J. Psychiatry 177, 435–446. https://doi.org/10.1176/appi.ajp.2019.19090915

Smith, A.H., Choi, K.S., Waters, A.C., Aloysi, A., Mayberg, H.S., Kopell, B.H., Figee, M., 2021. Replicable effects of deep brain stimulation for obsessive-compulsive disorder. Brain Stimul. 14, 1–3. https://doi.org/10.1016/j.brs.2020.10.016

Smith, S.M., Miller, K.L., Moeller, S., Xu, J., Auerbach, E.J., Woolrich, M.W., Beckmann, C.F., Jenkinson, M., Andersson, J., Glasser, M.F., Van Essen, D.C., Feinberg, D.A., Yacoub, E.S., Ugurbil, K., 2012. Temporally-independent functional modes of spontaneous brain activity. Proc. Natl. Acad. Sci. U. S. A. 109, 3131–3136. https://doi.org/10.1073/pnas.1121329109

Sriram, A., Foote, K.D., Oyama, G., Kwak, J., Zeilman, P.R., Okun, M.S., 2014. Brittle

dyskinesia following STN but not GPi deep brain stimulation. Tremor and Other Hyperkinetic Movements 4, 242. https://doi.org/10.7916/D8KS6PPR

Steigerwald, F., Müller, L., Johannes, S., Matthies, C., Volkmann, J., 2016. Directional deep brain stimulation of the subthalamic nucleus: A pilot study using a novel neurostimulation device. Mov. Disord. 31, 1240–1243. https://doi.org/10.1002/mds.26669

Strimbu, K., Tavel, J.A., 2010. What are biomarkers? Curr. Opin. HIV AIDS 5, 463–466. https://doi.org/10.1097/COH.0b013e32833ed177

Sullivan, C.R.P., Olsen, S., Widge, A.S., 2021. Deep brain stimulation for psychiatric disorders: From focal brain targets to cognitive networks. Neuroimage 225, 117515. https://doi.org/10.1016/j.neuroimage.2020.117515

Swanson, L.W., 2003. Brain architecture: Understanding the basic plan, 1st ed. Oxford University Press, New York, NY.

Swanson, L.W., 2000. Cerebral hemisphere regulation of motivated behavior. Brain Res. 886, 113–164. https://doi.org/10.1016/s0006-8993(00)02905-x

Sweet, J.A., Thyagaraj, S., Chen, Z., Tatsuoka, C., Staudt, M.D., Calabrese, J.R., Miller, J.P., Gao, K., McIntyre, C.C., 2020. Connectivity-based identification of a potential neurosurgical target for mood disorders. J. Psychiatr. Res. 125, 113–120. https://doi.org/10.1016/j.jpsychires.2020.03.011

Synofzik, M., Fins, J.J., Schlaepfer, T.E., 2012. A neuromodulation experience registry for deep brain stimulation studies in psychiatric research: Rationale and recommendations for implementation. Brain Stimul. 5, 653–655. https://doi.org/10.1016/j.brs.2011.10.003

Thorsen, A.L., Kvale, G., Hansen, B., van den Heuvel, O.A., 2018. Symptom dimensions in obsessive-compulsive disorder as predictors of neurobiology and treatment response. Curr. Treat. Options Psychiatry 5, 182–194. https://doi.org/10.1007/s40501-018-0142-4

Treu, S., Strange, B., Oxenford, S., Neumann, W.-J., Kühn, A.A., Li, N., Horn, A., 2020. Deep brain stimulation: Imaging on a group level. Neuroimage 219, 117018. https://doi.org/10.1016/j.neuroimage.2020.117018

Tsuboi, T., Wong, J.K., Eisinger, R.S., Okromelidze, L., Burns, M.R., Ramirez-Zamora, A., Almeida, L., Wagle Shukla, A., Foote, K.D., Okun, M.S., Grewal, S.S., Middlebrooks, E.H., 2021. Comparative connectivity correlates of dystonic and essential tremor deep brain stimulation. Brain 144, 1774–1786. https://doi.org/10.1093/brain/awab074

Tyagi, H., Apergis-Schoute, A.M., Akram, H., Foltynie, T., Limousin, P., Drummond, L.M., Fineberg, N.A., Matthews, K., Jahanshahi, M., Robbins, T.W., Sahakian, B.J., Zrinzo, L., Hariz, M., Joyce, E.M., 2019. A randomized trial directly comparing ventral capsule and anteromedial subthalamic nucleus stimulation in obsessive-compulsive disorder: Clinical and imaging evidence for dissociable effects. Biol. Psychiatry 85, 726–734. https://doi.org/10.1016/j.biopsych.2019.01.017

van den Heuvel, O.A., van Wingen, G., Soriano-Mas, C., Alonso, P., Chamberlain, S.R., Nakamae, T., Denys, D., Goudriaan, A.E., Veltman, D.J., 2016. Brain circuitry of compulsivity. Eur. Neuropsychopharmacol. 26, 810–827. https://doi.org/10.1016/j.euroneuro.2015.12.005

van der Vlis, T.A.M.B., Ackermans, L., Mulders, A.E.P., Vrij, C.A., Schruers, K., Temel, Y., Duits, A., Leentjens, A.F.G., 2021. Ventral capsule/ventral striatum stimulation in obsessive-compulsive disorder: Toward a unified connectomic target for deep brain stimulation? Neuromodulation 24, 316–323. https://doi.org/10.1111/ner.13339

van Eeden, W.A., van Hemert, A.M., Carlier, I.V.E., Penninx, B.W., Giltay, E.J., 2019. Severity, course trajectory, and within-person variability of individual symptoms in patients with major depressive disorder. Acta Psychiatr. Scand. 139, 194–205. https://doi.org/10.1111/acps.12987

Vanegas-Arroyave, N., Lauro, P.M., Huang, L., Hallett, M., Horovitz, S.G., Zaghloul, K.A., Lungu, C., 2016. Tractography patterns of subthalamic nucleus deep brain stimulation. Brain 139, 1200–1210. https://doi.org/10.1093/brain/aww020

Vedam-Mai, V., Deisseroth, K., Giordano, J., Lazaro-Munoz, G., Chiong, W., Suthana, N., Langevin, J.-P., Gill, J., Goodman, W., Provenza, N.R., Halpern, C.H., Shivacharan, R.S., Cunningham, T.N., Sheth, S.A., Pouratian, N., Scangos, K.W., Mayberg, H.S.,

Horn, A., Johnson, K.A., Butson, C.R., Gilron, R., de Hemptinne, C., Wilt, R., Yaroshinsky, M., Little, S., Starr, P., Worrell, G., Shirvalkar, P., Chang, E., Volkmann, J., Muthuraman, M., Groppa, S., Kühn, A.A., Li, L., Johnson, M., Otto, K.J., Raike, R., Goetz, S., Wu, C., Silburn, P., Cheeran, B., Pathak, Y.J., Malekmohammadi, M., Gunduz, A., Wong, J.K., Cernera, S., Wagle Shukla, A., Ramirez-Zamora, A., Deeb, W., Patterson, A., Foote, K.D., Okun, M.S., 2021. Proceedings of the Eighth Annual Deep Brain Stimulation Think Tank: Advances in optogenetics, ethical issues affecting DBS research, neuromodulatory approaches for depression, adaptive neurostimulation, and emerging DBS technologies. Front. Hum. Neurosci. 15, 644593. https://doi.org/10.3389/fnhum.2021.644593

Veerakumar, A., Berton, O., 2015a. Cellular mechanisms of deep brain stimulation: Activity-dependent focal circuit reprogramming? Curr Opin Behav Sci 4, 48–55. https://doi.org/10.1016/j.cobeha.2015.02.004

Veerakumar, A., Berton, O., 2015b. Cellular mechanisms of deep brain stimulation: Activity-dependent focal circuit reprogramming? Curr. Opin. Behav. Sci. 4, 48–55. https://doi.org/10.1016/j.cobeha.2015.02.004

Vergunst, F.K., Fekadu, A., Wooderson, S.C., Tunnard, C.S., Rane, L.J., Markopoulou, K., Cleare, A.J., 2013. Longitudinal course of symptom severity and fluctuation in patients with treatment-resistant unipolar and bipolar depression. Psychiatry Res. 207, 143–149. https://doi.org/10.1016/j.psychres.2013.03.022

Volkmann, J., Daniels, C., Witt, K., 2010. Neuropsychiatric effects of subthalamic neurostimulation in Parkinson disease. Nat. Rev. Neurol. 6, 487–498. https://doi.org/10.1038/nrneurol.2010.111

Wager, T.D., Woo, C.-W., 2017. Imaging biomarkers and biotypes for depression. Nat. Med. 23, 16–17. https://doi.org/10.1038/nm.4264

Wang, Q., Akram, H., Muthuraman, M., Gonzalez-Escamilla, G., Sheth, S.A., Oxenford, S., Yeh, F.-C., Groppa, S., Vanegas-Arroyave, N., Zrinzo, L., Li, N., Kühn, A., Horn, A., 2021. Normative vs. patient-specific brain connectivity in deep brain stimulation.

Neuroimage 224, 117307. https://doi.org/10.1016/j.neuroimage.2020.117307

Wang, Y., Zhang, C., Zhang, Y., Gong, H., Li, J., Jin, H., Li, D., Liu, D., Sun, B., 2020. Habenula deep brain stimulation for intractable schizophrenia: A pilot study. Neurosurg. Focus FOC 49, E9. https://doi.org/10.3171/2020.4.FOCUS20174

Waters, A.C., Veerakumar, A., Choi, K.S., Howell, B., Tiruvadi, V., Bijanki, K.R., Crowell, A., Riva-Posse, P., Mayberg, H.S., 2018. Test–retest reliability of a stimulation-locked evoked response to deep brain stimulation in subcallosal cingulate for treatment resistant depression. Hum. Brain Mapp. 39, 4844–4856. https://doi.org/10.1002/hbm.24327

Weigand, A., Edelkraut, L., Conrad, M., Grimm, S., Bajbouj, M., 2021. Light-dependent effects of prefrontal rTMS on emotional working memory. Brain Sci. 11, 446. https://doi.org/10.3390/brainsci11040446

Weigand, A., Horn, A., Caballero, R., Cooke, D., Stern, A.P., Taylor, S.F., Press, D., Pascual-Leone, A., Fox, M.D., 2018. Prospective validation that subgenual connectivity predicts antidepressant efficacy of transcranial magnetic stimulation sites. Biol. Psychiatry 84, 28–37. https://doi.org/10.1016/j.biopsych.2017.10.028

Whitton, A.E., Treadway, M.T., Pizzagalli, D.A., 2015. Reward processing dysfunction in major depression, bipolar disorder and schizophrenia. Curr. Opin. Psychiatry 28, 7–12. https://doi.org/10.1097/YCO.0000000000000122

Williams, L.M., 2017. Defining biotypes for depression and anxiety based on large-scale circuit dysfunction: A theoretical review of the evidence and future directions for clinical translation. Depress. Anxiety 34, 9–24. https://doi.org/10.1002/da.22556

Yamamoto, T., Kobayashi, K., Kasai, M., Oshima, H., Fukaya, C., Katayama, Y., 2005. DBS therapy for the vegetative state and minimally conscious state. Acta Neurochir. Suppl. 93, 101–104. https://doi.org/10.1007/3-211-27577-0_17

Yan, H., Boutet, A., Mithani, K., Germann, J., Elias, G.J.B., Yau, I., Go, C., Kalia, S.K., Lozano, A.M., Fasano, A., Ibrahim, G.M., 2020. Aggressiveness after centromedian nucleus stimulation engages prefrontal thalamocortical circuitry. Brain Stimul. 13, 357–

359. https://doi.org/10.1016/j.brs.2019.10.023

Yeh, F.-C., 2020. Shape analysis of the human association pathways. Neuroimage 223. https://doi.org/10.1016/j.neuroimage.2020.117329

Yücel, M., Oldenhof, E., Ahmed, S.H., Belin, D., Billieux, J., Bowden-Jones, H., Carter, A., Chamberlain, S.R., Clark, L., Connor, J., Daglish, M., Dom, G., Dannon, P., Duka, T., Fernandez-Serrano, M.J., Field, M., Franken, I., Goldstein, R.Z., Gonzalez, R., Goudriaan, A.E., Grant, J.E., Gullo, M.J., Hester, R., Hodgins, D.C., Le Foll, B., Lee, R.S.C., Lingford-Hughes, A., Lorenzetti, V., Moeller, S.J., Munafò, M.R., Odlaug, B., Potenza, M.N., Segrave, R., Sjoerds, Z., Solowij, N., van den Brink, W., van Holst, R.J., Voon, V., Wiers, R., Fontenelle, L.F., Verdejo-Garcia, A., 2019. A transdiagnostic dimensional approach towards a neuropsychological assessment for addiction: An international Delphi consensus study. Addiction 114, 1095–1109. https://doi.org/10.1111/add.14424

Zhou, C., Zhang, H., Qin, Y., Tian, T., Xu, B., Chen, J., Zhou, X., Zeng, L., Fang, L., Qi, X., Lian, B., Wang, H., Hu, Z., Xie, P., 2018. A systematic review and meta-analysis of deep brain stimulation in treatment-resistant depression. Prog. Neuro-Psychopharmacology Biol. Psychiatry 82, 224–232. https://doi.org/10.1016/j.pnpbp.2017.11.012